\documentclass[revtx4]{emulateapj}
\usepackage{natbib}
\usepackage{abrev} %astronomical abbreviations
\usepackage{multirow}
\begin{document}

\title{The effect of models of the interstellar media on the central mass distribution of galaxies}
\author{C.~R.~Christensen}
\affil{Department of Astronomy, University of Arizona, 933 North Cherry Avenue, Rm. N204, Tucson, AZ 85721-0065, USA}
\author{F.~Governato and  T.~Quinn}
\affil{Department of Astronomy, University of Washington, Box 351580, Seattle, WA 98195, USA}
\author{A.~M.~Brooks}
\affil{Department of Astronomy, University of Wisconsin-Madison, 475 N. Charter St., Madison, WI 53706, USA}
\author{D.~B.~Fisher}
\affil{Department of Astronomy, University of Maryland, CSS Bldg., Rm. 1204, Stadium Dr., College Park, MD 20742-242, USA}
\author{S.~Shen}
\affil{epartment of Astronomy and Astrophysics, University of California, Santa Cruz, CA 95064, USA}
\author{ J.~McCleary}
\affil{Department of Physics, Brown University, Box Box 1843, 182 Hope St., Barus \& Holley, Providence, RI 02912, USA}
\author{ J.~Wadsley}
\affil{Department of Physics and Astronomy, McMaster University, Hamilton, ON, Canada}

\begin{abstract}
We compare the central mass distribution of galaxies simulated with three different models of the interstellar medium (ISM) with increasing complexity: primordial (H+He) cooling down to 10$^4$K, additional cooling via metal lines and to lower temperatures, and molecular hydrogen ($\Hmol$) with shielding of atomic and molecular hydrogen, in addition to metal line cooling.
The latter model includes a non-equilibrium calculation of the $\Hmol$ abundance, self-shielding of $\Hmol$, dust shielding of both HI and $\Hmol$, and $\Hmol$-based star formation efficiency.
In order to analyze the effect of these models while holding all other parameters constant, we follow the evolution of four field galaxies with V$_{peak} < $ 120 km/s to a redshift of zero using high-resolution Smoothed Particle Hydrodynamic simulations in a fully cosmological $\Lambda$CDM context.
The spiral galaxies produced in simulations with either primordial cooling or $\Hmol$ physics have  realistic, rising rotation curves.
In contrast, the simulations with metal line cooling and otherwise similar feedback and star formation produced galaxies with the peaked rotation curves typical of most previous $\Lambda$CDM simulations of spiral galaxies.
The less-massive bulges and non-peaked rotation curves in the galaxies simulated with primordial cooling or $\Hmol$ are linked to changes in the angular momentum distribution of the baryons.
These galaxies had smaller amounts of low-angular momentum baryons because of increased gas loss from stellar feedback.
When there is only primordial cooling, the star forming gas is hotter and the feedback-heated gas cools more slowly than when metal line cooling is included and so requires less energy to be expelled.
When $\Hmol$ is included, the accompanying shielding produces large amounts of clumpy, cold gas where $\Hmol$ forms. 
Consequentially, star formation takes place in cold, dense molecular gas where supernova feedback is more effective at driving outflows.
The higher feedback efficiency causes decreased low-angular momentum material and the formation of realistic rotation curves.
The inclusion of the $\Hmol$ model and the resulting increase in feedback enabled the creation of simulations with both a more physically-based cooling model and realistic structure within the central kiloparsec.
\end{abstract}

\section{Introduction}

The central question in  $\Lambda$-Cold Dark Matter ($\Lambda$CDM) galaxy formation is how the gas collected into dark matter (DM) halos was transformed into the stellar populations we observe in galaxies today.
To answer this question, we must understand how the interplay between the physical processes connected with star formation resulted in the observed stellar content of modern galaxies.
These processes include the rate at which gas cools, how it is transformed into stars, and the regulation of star formation by such things as stellar feedback.
Historically, it has been difficult for $\Lambda$CDM simulations to reproduced observed stellar bulges and the central regions of rotation curves.
In this work, we concentrate on how different models of the interstellar media (ISM) and star formation impact the amount of mass in the central region of the disk galaxies, an area particularly sensitive to changes in the angular momentum distribution (AMD).
We demonstrate how substantial redistribution of the angular momentum in simulated disk galaxies can be induced by changes to the ISM and, consequentially, to the star formation and supernova (SN) feedback.

Simulations offer the ideal format for studying the evolution of the non-linear processes that dominate galaxy formation over cosmic time, such as the formation of galactic disks.
Disks are created by preserving angular momentum from cosmic torques during the infall of gas \citep{white84, barnes87, Quinn92, maller02a, vitvitska}.
If the gas infall is smooth, high-angular momentum gas will form a similarly high-angular momentum disk \citep[e.g.][]{mo98, dalcanton97}.
In order to preserve the disk, large amounts of high-angular momentum material must be maintained throughout the galaxy's evolution.

Excess angular momentum loss has been shown through galaxy simulations to result from insufficient feedback; gas cools and forms stars too quickly and angular momentum is subsequently lost during mergers \citep{WhiteANDReese78, Dekel86, nb91, Efstathiou92, katz94, navarro94, thoul, quinn96, navarrosteinmetz97, Sommerlarsen99, gnedin00, Maller02, SommerLarsen03, vdb03, robertson04, Okamoto05, donghia06, Hoeft06, Governato08, Okamoto08, Scannapieco08, zavala08, ceverino09, dutton09b, keres09, Sales10}.
Further work has shown that low resolution also contributes to angular momentum loss through such means as two-body heating and artificial viscosity \citep{Thacker00, Sommerlarsen99, Mayer01,Governato04, Kaufmann07, Governato07, Naab07, Mayer08, piontek11}. 
Other simulations have suggested that disk survival could be aided by reducing star formation efficiency at high redshift \citep{Gnedin10}.
Recently, simulations have progressed in resolution and the modeling of feedback to the point that angular momentum is conserved and disks of the appropriate size that lie along the Tully-Fisher relation are produced even with active merger histories \citep{Governato04, Robertson06, Scannapieco08, hopkins09, Agertz10, Brooks11, Brook11b, Guedes11}.

Importantly, though, even if the initial angular momentum imparted by tidal torques can be maintained, the resulting galaxies do not match the morphologies of observed galaxies.  
Although observed disk sizes can be reproduced \citep{Fall80, dalcanton97, mo98, Brooks11}, the resulting bulges predicted in $\Lambda$CDM are much too large compared to observed bulges \citep{Bullock01, vandenBosch01a, vandenBosch01b, Binney01, Maller02, maller02a, vdb02, donghia06, dn07,Stinson10}. 
This disagreement is true across all disk galaxy masses, not only at the bulgeless dwarf galaxy scale at which the problem is most severe \citep{dutton09disks}.  
The discrepancy in the amount of low-angular momentum material argues that some process {\em must alter the AMD of disk galaxies by preferentially removing the low-angular momentum gas} that would naturally builds up the central regions and bulges of galaxies.

The most sensitive area in the disk to the amount of low-angular momentum material is the central 1 kpc.
In this work, therefore, we focus on how the problem of overly large bulges in simulations is affected by different gas cooling models.  
The study of the central regions of galaxies in relation to angular momentum has only recently become attainable because of computational demands.
Cosmological simulations are necessary to follow the detailed evolution of the AMD.
Such simulations, however, are computationally expensive as they require force resolutions small enough to resolve the inner $\sim$1 kpc, i.e. $< 200$ pc so that the inner 1 kpc contains at least five resolution units.

A few such simulations that resolve the internal structure of disks have demonstrated recently that low-angular momentum gas can be preferentially removed from the galaxy \citep{Governato10, Brook11b, Brook11a, Maccio12}.
In \citet{Governato10} and \citet{Brook11b}, these outflows led to bulgeless disk galaxies with realistic rotation curves \citep{Oh11b}.  
\cite{Guedes11} outlined the creation of a late-type spiral galaxy with a realistic bulge-to-total disk mass ratio, created by centralized gas loss. 
\citet{Brook11a} demonstrated that the outflows created in these high resolution simulations naturally drive low-angular momentum gas from the galaxy.  
In \citet{Stinson12}, the addition of early stellar feedback led to both stronger outflows and galaxies with more realistic bulges.
Results from the Aquila comparison project also support the idea that increased feedback is connected to flatter rotation curves in Milky Way-mass galaxies.
In this project, flat rotation curves were correlated with stronger galactic winds \citep{scannapieco11}.
These same outflows that remove low-angular momentum gas create massive, metal-enriched halos around galaxies that are similar to  those inferred from absorption-line spectra \citep{Stinson12, Shen13}.

The preferential removal of low-angular momentum material is key to correctly simulating the central mass distributions in galaxies.
In order for this removal to take place, it is crucial to realize the 3-D structure of the ISM at the scale of star formation.
When star formation in simulations is limited to regions with densities comparable to those of observed molecular clouds, gas is able to collect in clumps with densities $ \geq$100 amu/cc.  
When the massive stars in these clumps produce SN, the feedback is highly concentrated, leading to expanding bubbles of hot gas that puncture the disk and naturally drive galactic outflows.
Importantly, the ability to resolve these dense star forming clumps allows the gas to reach comparable densities to the DM in the central regions of galaxies.  
Due to these comparable densities, sudden gas loss caused by SN feedback can flatten the gravitational potential, leading to an expansion of the DM orbits and a flattening of the DM density profile \citep[i.e., a shift from a more cuspy profile to a more cored profile][]{Pontzen11}.  
Hence, both the creation of outflows that remove low-angular momentum gas and the creation of DM cores rely strongly on star formation happening in regions of high density gas.

A natural and physically-motivated method for ensuring that star formation and feedback take place in high density environments in the simulations is to connect star formation to the radiation-shielded areas of the ISM where $\Hmol$ is abundant \citep{RobertsonKravtsov08}.
In observed galaxies, star formation is strongly correlated with molecular gas \citep{Rownd99, Murgia02, Heyer04, Gao04, Kennicutt07,Blanc09,Warren10,Bigiel10,Schruba11,Bolatto11}.
Basing star formation on the $\Hmol$ abundance also introduces additional dependencies, such as on metallicity.
Because of this metallicity dependency, connecting star formation to $\Hmol$ has been shown to lower the star formation efficiency in simulations of the low-metallicity, high-redshift Universe \citep{Gnedin10, Gnedin11, Kuhlen11, Krumholz11a}.
Reducing high-redshift star formation may lower the amount of angular momentum loss during mergers \citep{hopkins09}.
However, reducing early star formation cannot be responsible for reducing the amount of low-angular momentum material below the initial AMD of the gas.

As of yet, most simulations that have incorporated $\Hmol$-based star formation have only used inefficient thermal SN feedback, making them unable to drive outflows that redistribute the angular momentum of the disk gas.
\citet{Christensen12}, however, demonstrated that using both $\Hmol$-based star formation and efficient SN feedback to simulate a dwarf galaxy changed the distribution of star formation and the temperature and density distribution of the gas.
In this paper, the addition of shielding with the $\Hmol$ physics was largely responsible for the increased amount of dense, cold gas.
Furthermore, an $\Hmol$-based star formation recipe limited star formation to such dense gas.
This ability of $\Hmol$ to influence the location and environment of star formation suggests it could also affect the redistribution of angular momentum through feedback, making it a promising tool to better match observational trends in galaxies.

The possible effect of the inclusion of shielded $\Hmol$ gas on feedback efficiency reflects a more general dependency of star formation and the feedback efficiency on the ISM model.
The rates of cooling and heating naturally affect the mass, distribution, and clumpiness of the gas disk in galaxies.
The ability of gas to accrete onto the disk of the galaxy regulates the amount of gas in the disk and, consequentially, the amount of gas available for star formation \citep{ChoiNagamine09,Schaye09,vdv11}.  
Greater amounts of cooling can also make it more difficult to expel gas through feedback.
On the other hand, additional low-temperature cooling in the disk leads to the formation of more clumps, potentially increasing the amount of outflowing gas and further altering the AMD.
Other changes to the distribution of star formation and the pressure, temperature, or density of the gas have the potential to further alter the efficiency of the feedback model.

The galaxies with realistic disk scale lengths produced by \citet{Guedes11} and \citet{Governato10} were simulated without full metal line cooling.
This raises  the question of how they would fare with a more realistic cooling model.
\citet{Scannapieco08} and \citet{piontek11} allowed for cooling above $10^4$ K from metal lines but produced galaxies with unphysically large bulges.
\citet{Brook11b, Brook11a}, and \citet{Maccio12} also included metal line cooling in their simulations but were able to produce realistic bulges only through the addition of an extra source of energy from young stars.  
Of the simulations that produced flat rotation curves in the Aquila project without AGN feedback \citep{scannapieco11}, both G3-TO \citep{Okamoto10} and G3-CS \citep{Scannapieco05, Scannapieco06} included metal-based cooling above $10^4$ and a subgrid-model for the multi-phase ISM.
In the latter simulation, this subgid model of the ISM is specifically shown to increase the efficiency of SN feedback.
These simulations indicate that metal-line cooling in the absence of enhanced stellar feedback models results in an increased amount of central material in galaxies.

\begin{figure*}
 \includegraphics[width=1.0\textwidth]{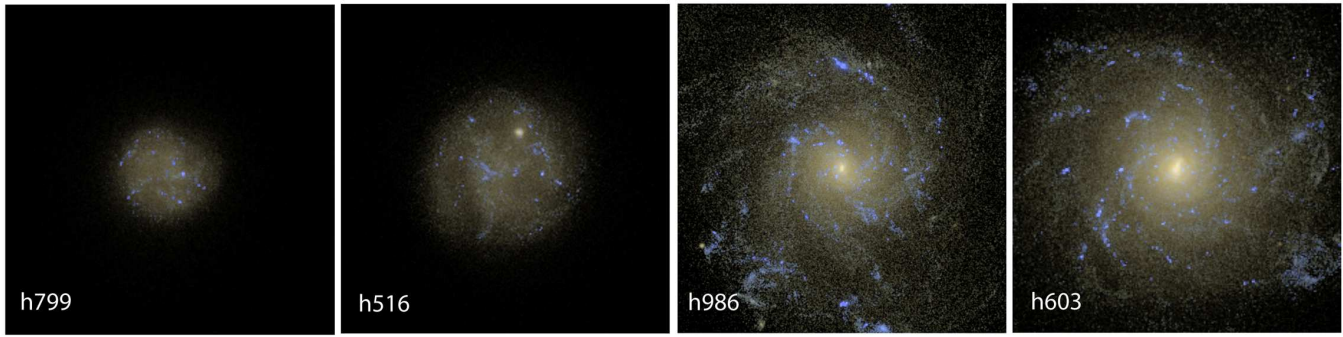} 
 \caption[Simulated observations of the four galaxies]
 {Mock-observations of the two dwarf (left panels) and spiral galaxies (right panels) used in this paper simulated with metal line cooling and $\Hmol$ physics.
The left (dwarf) frames are 12 kpc across and the right (spiral) frames are 24 kpc across.
These SDSS multicolor ({\em r}, {\em g}, and {\em i}) images of the galaxies at z$=$0 were generated with {\sc Sunrise}.
}
\label{fig:morphology}
\end{figure*}

Ultimately, simulations should move to more detailed and accurate modeling of the physics of galaxy formation and, in this instance of the ISM.
For example, simulations without a model for cooling from metal lines will not be able to correctly simulate the cooling of metal-enriched halo material onto galactic disks.
The addition of more accurate physics, such as metal-line cooling and the modeling of the shielded ISM have potentially far-reaching consequences.
As we move to these more detailed models, it is critical to understand how they impact other processes, such as feedback, so that the various models will act in concert.
With that goal in mind, we take pains to compare the effects of these three ISM models while holding all other variables constant.
Such an analysis will both clarify our understanding of the link between ISM models and other processes and inform the future tuning of other simulation parameters to the different ISM models.

Here, we analyze the effect of different ISM models on star formation and the AMD in simulations of galaxies.
To do this, we simulated a small set of dwarf and spiral galaxies of different masses to z = 0 using three different models of the ISM: primordial H+He cooling down only to $10^4$ K, the addition of metal line cooling, and both metal line cooling and cooling from shielded $\Hmol$ gas.
Covering multiple galaxy masses and different observable quantities is necessary to constrain the multivariate problem of the interaction between the ISM, star formation and feedback. 
We expect that the different ISM models will have less of an effect on the dwarf galaxies because of their smaller metallicities and lower $\Hmol$ content.  
In these low-metallicity galaxies the inclusion of the additional cooling from metals or shielding will produce smaller changes to the amount and distribution of cold gas.
As the galaxy mass increases, though, they will become more metal rich with larger amounts of cold, shielded gas.
The choice of ISM model will, consequentially, have a greater effect on the structure of the ISM and the resulting efficiency of SN feedback.
In order to compare our simulations to observations, we generated and analyzed both optical and IR images as well as HI velocity cubes.
We then examined several different metrics, including the rotation curves, the structure of the ISM, the star formation environment, and the efficiency of feedback to determine the effect of the ISM models.

The paper is organized as follows: \S\ref{sec:method5} discusses the simulations; \S\ref{sec:globalProperties} describes the mass distributions of the galaxies; \S\ref{sec:ISM} describes differences in the ISM caused by the introduction of $\Hmol$; \S\ref{sec:decomp} analyzes star formation in the galaxies, and \S\ref{sec:snfb} compares the feedback efficiency.
The results are then discussed in \S\ref{sec:discuss} and \S\ref{sec:res5} is the conclusion.

%--------------------------------- Methods ---------------------------
\section{Methods: Description of the Simulation}\label{sec:method5}
The simulations described here are part of a new set of high-resolution simulations aimed at studying the formation of field galaxies.
Our sample of halos includes two dwarf galaxies (virial masses of a few $10^{10}$ M$_\odot$) and two spiral galaxies (virial masses of several times $10^{11}$ M$_\odot$).
The properties of the simulated galaxies are listed in Table 1.

The simulations were computed in a WMAP3 cosmology \citep{Spergel07}: $\Omega_0$=0.24, $\Omega_{baryon}$=0.04, $\Lambda$=0.76, h=0.73, $\sigma_8$=0.77. 
Our halo sample was selected from lower resolution cosmological volumes and then re-simulated at much higher resolution using the `zoom-in' technique \citep{Katz92}.
Adopting the zoom-in approach enabled us to have a significant number of high-resolution particles while still following the surrounding large-scale structure.
There were more than one million DM particles within the virial radius of each galaxy at z=0.
For the dwarf (spiral) galaxies the DM, gas, and star particle masses (at the time of formation) were respectively: 1.6 (13)$\times$10$^4$, 3.3 (27.0)$\times$10$^3$, and 1.0 (8.0)$\times$10$^3$M$_{\odot}$.  
The force spline softening was 87 pc (170 pc) for the dwarf (spiral) galaxies. 
The minimum smoothing length for gas particles was 0.1 times the force softening. 
This resolution was sufficient to follow the formation of star forming regions as small as 10$^4$M$_{\odot}$.
We further discuss resolution in relation to star formation in \S\ref{sec:SF}.

%Description of Gasoline
The simulations were performed with the $N$-body SPH code {\sc GASOLINE} \citep{Wadsley04} with a force accuracy criterion of $\theta$ = 0.725, a time step accuracy of $\eta$=0.195 and a Courant condition of $\eta_C$=0.4.  
The simulations included a cosmic UV field modeled following an updated \cite{Haardt96}, which partially suppresses the collapse of baryons into the smallest halos \citep{Hoeft06, Governato07}.
Our adopted star formation and SN schemes have been described in detail in \cite{Stinson06} and \cite{Governato07}.
Briefly, the ``blastwave'' feedback scheme is implemented by releasing thermal energy from SN into gas surrounding young star particles.  
The heated gas particles have their cooling shut off for a time equal to the momentum-conserving phase of the SN blastwave.
This time is typically a few million years and is a function of the local density and temperature of the gas and the amount of energy injected and the cooling. 
In these simulations, we assume that all of the canonical $10^{51}$ ergs released per supernova ($E_{SN}$) to be transferred to the ISM.
Metals are injected into the gas by type I and II SN following \citet{Raiteri96} and distributed across the smoothing sphere. 
{Metals are also injected by stellar winds; our model for mass loss follows \citet{Weidemann87} and assumes the metallicity is that of the stellar particle.
Metal diffusion \citep{Shen10} is responsible for distributing the metals throughout the galaxy.

%Star formation
As described in the following subsection, star formation was limited to cold gas in dense regions, with criteria depending on the ISM model (see \S\ref{sec:SF}). 
Only three main free parameters in the SN feedback scheme (the star formation efficiency, density threshold for star formation, and the fraction of SN energy coupled to the ISM) were fixed to reproduce the properties of present day galaxies over a range of masses \citep{Governato07}. 
Without further adjustments, this scheme has been shown to reproduce the relation between metallicity and stellar mass \citep{Brooks07, Maiolino08} and the abundance of Damped Lyman $\alpha$ (DLA) systems at z=3 \citep{Pontzen08}.

\begin{figure*}
 \includegraphics[width=1.0\textwidth]{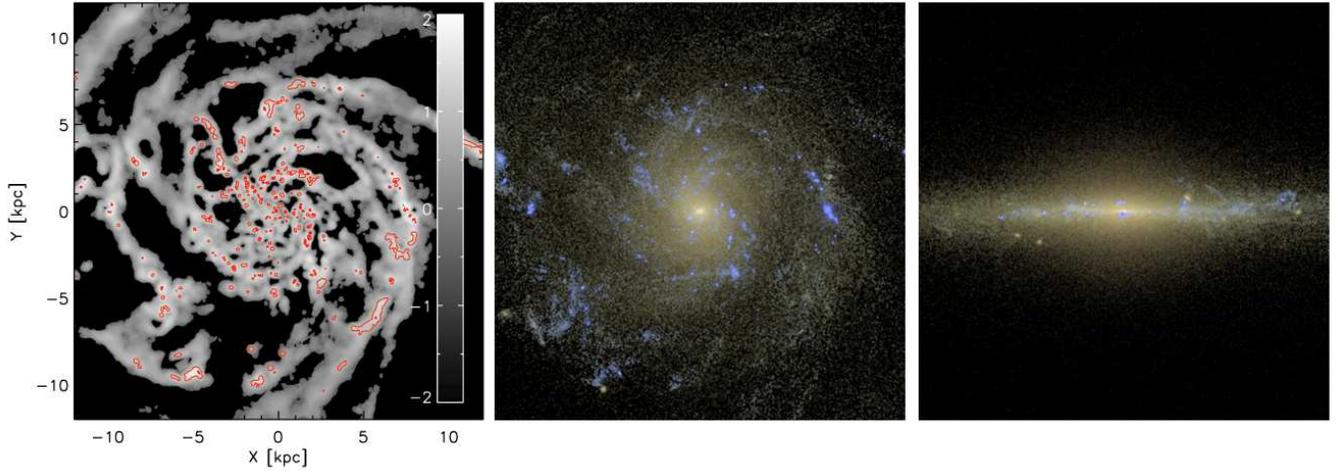} 
 \caption[Gas surface density and simulated observations of an L$^*$ galaxy]
 { Gas surface density and simulated observations of a spiral galaxy (h986) simulated with metal line cooling and $\Hmol$ physics.
 Each frame is 24 kpc across.
Left: HI surface density (gray scale) overlaid with red contours indicating the $\Hmol$ surface density.
The red contours are shown for the following levels: log $\Sigma_{\Hmol}$ = -2, 0, and 2 amu/cc.  
Middle and right: SDSS multicolor ({\em r}, {\em g}, and {\em i}) images of the same galaxy at z$=$0 generated with SUNRISE in which the galaxy is oriented face on and edge on, respectively.
}
\label{fig:pretty_pic}
\end{figure*}

\subsection{Models of the ISM}
%Introduce different models of ISM
We computed the simulations using three different ISM models of increasing complexity and their corresponding star formation recipes.
In these models, the different heating and cooling rates and the presence or lack of shielding resulted in varying temperature and density distributions.
Because of the different structure of the ISM, the interaction between the ISM, star formation, and feedback was also altered.
Here we outline each of the models.

%What is the same across all models
All the considered models of the ISM include a calculation of the non-equilibrium ion abundances of H and He.
The redshift-dependent UV background radiation was responsible for both gas heating and photoionization.
Gas cooling was calculated based on collisional ionization \citep{Abel97}, radiative recombination \citep{Black81, VernerANDFerland96}, photoionization, bremsstrahlung, and H and He line cooling \citep{Cen92}.

%Primordial cooling
In our simplest ISM model, the heating and cooling of the gas is due only to the above reactions in the neutral and ionized H and He.
For the purposes of this paper, we label this the ``primordial'' ISM model.
Because of the lack of low-temperature coolants in this model, cooling is inefficient below $10^4$ K.
Gas is, therefore, unable to reach lower temperatures and the formation of high density regions is suppressed by the extra pressure support.

%Metal Lines
Additional cooling from metal lines \citep{Shen10} is included in our second ISM model (labeled ``metal'' throughout the paper).
With the addition of metal line cooling, gas is able to cool more efficiently and to temperatures below $10^4$ K.
These metal line cooling rates were computed for different gas temperatures, densities, metallicities, and amounts of UV background assuming ionization equilibrium and optically thin gas using CLOUDY \citep[version 07.02;][]{Ferland98}. 
The metal line cooling rates were introduced to model the intergalactic medium.
In this regime it is safe to assume that the gas is optically thin and that the cosmological UV background dominates the radiation field.
In the ISM, however, this assumption can result in an underestimate of cooling rates.
In the ISM, the lack of an interstellar radiation field beyond the UV background results in substantially less CII \citep{Christensen12}, the most important coolant in spiral galaxies.
Furthermore, the lack of shielding from photoionizing radiation prevents the additional formation of cold gas.

%H2 model
Our most sophisticated ISM model (labeled as the ``$\Hmol$'' model) includes the non-equilibrium formation of $\Hmol$ \citep{Christensen12} in addition to metal line cooling.
Molecular hydrogen forms primarily on dust grains and is dissociated by Lyman-Werner radiation from young stars.
It is, therefore, limited to dense regions of the ISM where the gas is shielded from Lyman-Werner radiation.
Our model follows the hydrogen chemical network for each gas particle, assuming that the dust fraction is proportional to the metallicity and calculating the incident Lyman-Werner radiation based on the emission from nearby star particles.
Our rate of $\Hmol$ formation on dust grains is proportional to gas density and metallicity and based on the observationally motivated equation from \citet{Wolfire08}.
As in \citet{Gnedin09}, we model the unresolved density structure of gas particles when calculating the formation rate using a sub-grid clumpiness factor ($C_p$ = 10, as in \citet{Christensen12}).
In addition to $\Hmol$ formation on dust grains, we model the gas-phase formation of $\Hmol$ using the minimal model for $\Hmol$ formation via H$^-$ from \citet{Abel97}.
Because of computational expense, we do not include full radiative transfer when calculating the Lyman-Werner radiation or the shielding.
We model the local flux of the Lyman-Werner radiation at the position of each gas particle based off the emission from nearby stellar particles, using Starburst99 \citep{Leitherer99} to calculate the emission from a simple stellar population of the same age and mass as the star particle.
We assume optically thin gas outside of the molecular clouds (although, as will be discussed, we include shielding for individual particles) and use the gravitational tree structure to determine the average Lyman-Werner flux within a given cell.
Our model includes self-shielding and shielding by dust to protect $\Hmol$ from photo-dissociation; dust shielding is also included when calculating the rates of HI photoionization and photoheating.
Following the work of \citet{Draine96}, \citet{Glover07} and \citet{Gnedin09}, we use a phenomenological model for the shielding based on the surface density of each gas particle.
In order to calculate the surface density for each particle, we assume a column length equal to the smoothing length.

The inclusion of shielded HI and $\Hmol$ strongly decreases the amount of heating from photo-dissociation and ionization and results in the formation of a cold (100 K) ISM.
Some additional low-temperature cooling is provided through $\Hmol$ collisions \citep{Gnedin11}.
While the more accurate calculation of CII and other low-temperature metal coolants would increase the cooling rates, the formation of cold gas is primarily dependent on the presence of shielding.
This model for the ISM is, therefore, an important step toward the accurate modeling of the cold, shielded phase of the ISM where star formation takes place.

\subsection{Star Formation}
\label{sec:SF}
%Base Star Formation Model
As described in  \cite{Stinson06}, star formation in these simulations occurs stochastically in gas that is both sufficiently dense and cool.
For a gas particle to be eligible to spawn a star particle in this model, it must have a temperature less than $T_{max}$ and a  density greater than $n_{min}$.
For eligible gas particles, the probability, $p$, of a star particle forming is a function of the dynamical time, $t_{form}$,

\begin{equation}
p = \frac{m_{gas}}{m_{star}}(1 - e^{-c^* \Delta t /t_{form}})
\end{equation}
where $m_{gas}$ is the mass of the gas particle, $m_{star}$ is the initial mass of the potential star particle, and $c^*$ is a star forming efficiency factor.

%Star Formation in primordial and metal model
We used different star formation parameters ($c^*$, $T_{max}$, $n_{min}$) based on the ISM model.
For both the primordial and metal ISM models,  $c^* = 0.1$.
We set $T_{max} = 10^4$K and $n_{min} = $10 (100) amu cm$^{-3}$ for the spiral (dwarf) galaxies in order to select for the gas corresponding to the cold ISM in these simulations.
The high value for $T_{max}$ reflects the inability of gas in the primordial ISM model or zero metallicity gas in the metal line cooling ISM model to cool below $10^4$ K.
Star formation is, therefore, allowed to occur in gas substantially hotter than the $\sim$100K gas of molecular clouds.
These high density thresholds for star formation are similar for those chosen for previous simulations of spiral galaxies \citet{Guedes11} and dwarf galaxies \citet{Governato10}.
The high thresholds are intended to limit star formation to gas with densities similar to molecular clouds.
In spiral galaxies, this transition to star forming molecular clouds occurs at lower densities than in dwarf galaxies and the differences in $n_{min}$ reflect the differences in transition densities.
However, it is clear that the tuning of a parameter for a final galactic mass in a galaxy formation simulation is inherently problematic, as multiple mass galaxies form within the same box and the galaxy mass changes with time.
One of the motivations of tying star formation of $\Hmol$ is to physically represent this change in threshold.

%Star formation in H2 model
When using the $\Hmol$ model, we incorporated the abundance of $\Hmol$ into the star formation calculation since star formation is observationally linked to $\Hmol$ \citep[e.g.][]{Bigiel10}.
We modeled this connection between star formation and $\Hmol$ by making the star formation efficiency a function of the local $\Hmol$ abundance: $c^* = 0.1 X_{\Hmol}$, where $X_{\Hmol}$ is the fraction of hydrogen in the form of $\Hmol$.
This functionally eliminates the need for a high density threshold to restrict star formation to molecular clouds.
We, therefore, set the density threshold to the arbitrarily low value of $n_{min} = 0.1$ amu.
We further limited star formation in this recipe to gas that has undergone molecular cooling by setting $T_{max}  = 1000$ K.
We tested the sensitivity of star formation to both $n_{min}$ and $T_{max} $ when using $\Hmol$ and found that changes of a factor of five to either threshold did not significantly alter the star formation.
With this star formation recipe, we naturally limit star formation to regions that correspond best to those in observed galaxies.
Both the inclusion of $\Hmol$ and the $\Hmol$-based star formation recipe were tested on a cosmological dwarf galaxy in \citet{Christensen12} and found to produce a bluer dwarf galaxy at z = 0 with more extended star formation and clumpier gas.

%Background on Jeans instability
The high density thresholds for star formation in combination with the low temperature cooling allow gas particles to become Jeans unstable.
While this can occur with metal line cooling, the formation of cold gas when shielding is included mean that it is far more prevalent in the $\Hmol$ ISM model.
\citet{BateANDBurkert97} showed that when the Jeans length or mass is unresolved artificial fragmentation can occur.
At this point star formation becomes dependent on the specifics of the star formation implementation.
Many simulators have addressed this by requiring that Jeans instability never occur, either by not including low temperature cooling or adding a pressure floor.
Still others made Jeans instability itself the criteria for star formation.
We allow Jeans instability to occur with the understanding that gas particles that are Jeans unstable will soon form stars and that the highly efficient SN feedback will heat the surrounding gas, preventing further collapse.
While this method does make the star formation dependent on the implementation,  Jeans instability is a natural occurrence and it would be equally artificial to prevent it.
We see no evidence of artificial collapse in our simulations.
We caution, however, that these simulations should not be used to study phenomena occurring below our resolution limit, such as the mass-spectrum of molecular clouds.

\subsection[mockO]{Properties of individual galaxies}
\label{sec:mocO}

\begin{table*}
\begin{center}
\begin{tabular}{l|cccccccccccccc}                                        
&				& Virial Mass			& Gas Mass				& Cold Gas		& Stellar 				& $V_f$	         & Optical	& SFR$_{\mathrm{H}_\alpha}$	& \% Gas	\\
&				&  					& in $R_{vir}$  				&Mass 			& Mass 				& 			& Radii	&						& Lost	\\ 
&				& [$\Msun$]			& [$\Msun$] 				& [$\Msun$]  		& [$\Msun$]  			& [km/s] 		& [kpc]	& [M$_{\odot}$/yr] 			& 		\\
&				& (1)					& (2)						& (3)				& (4)					& (5)			& (6)		& (7)						& (8) \\\hline \hline

\multirow{3}{*}{{\bf h516$^{1 - 8}$}}
&Primordial		& $3.9\times 10^{10}$	& $2.6\times 10^{9}$			& $1.0\times 10^{9}$& $2.6\times 10^{8}$		& 61			& 1.2		& 0.004				         & 10		\\
&Metals			& $3.6\times 10^{10}	$	& $1.7\times 10^{9}$			& $3.4\times 10^{8}$& $2.1\times 10^{8}$		& 60			& 1.4		& 0.005					& 10		\\
&$\Hmol$ 	   	& $3.8\times 10^{10}	$	& $2.3\times 10^{9}$			& $3.9\times 10^{8}$& $2.5\times 10^{8}$		& 65			& 1.9		& 0.011					& 12		\\
\hline
\multirow{3}{*}{{\bf h799$^{1,2,6,7}$}}
&Primordial		& $2.3\times 10^{10}	$	& $1.3\times 10^{9}$			& $5.6\times 10^{8}$	& $1.8\times 10^{8}$		& 58			& 0.8		& 0.001					& 10		\\
&Metals			& $2.4\times 10^{10}	$	& $1.5\times 10^{9}$			& $3.8\times 10^{8}$	& $1.3\times 10^{8}$		& 60			& 0.8		&  -						& 7		\\
&$\Hmol$  		& $2.4\times 10^{10}$	& $1.4\times 10^{9}$			& $2.1\times 10^{8}$	& $1.4\times 10^{8}$		& 54			& 1.3		& 0.007					& 15		\\
\hline
%-------------------------------------------------------------------------------------
%{\bf h603} 		& 					&						&				&					&			&						&\\  
\multirow{3}{*}{{\bf h603$^{1,2,7}$}}
&Primordial		& $3.1\times 10^{11}	$	& $2.8\times 10^{10}$		& $2.0\times 10^{9}$	& $5.5\times 10^{9}$		& 104		& 5.5		& 0.068					& 15		\\
&Metals			& $3.0\times 10^{11}	$	& $2.6\times 10^{10}$		& $2.3\times 10^{9}$	& $1.3\times 10^{10}$	& 132		& 7.1		& 0.182					& 8		\\
&$\Hmol$ 		& $3.4\times 10^{11}$	& $3.1\times 10^{10}	$		& $3.4\times 10^{9}$& $7.8\times 10^{9}$		& 111		& 7.2		& 0.389					& 10		\\
\hline
%------------------------------------------------------------------------------------
%{\bf h986} 		& 					&						&				&					&			&						&\\  
\multirow{3}{*}{{\bf h986$^{1,2,7}$}}
&Primordial		& $1.7\times 10^{11}$	& $1.4\times 10^{10}$		& $1.2\times 10^{9}$& $4.0\times 10^{9}$		& 97			& 5.7		& 0.023					& 16		\\ 
&Metals			& $1.8\times 10^{11}$	& $1.5\times 10^{10}$		& $1.5\times 10^{9}$& $1.1\times 10^{10}$	& 131		& 7.3		& 0.354					& 7		\\
&$\Hmol$  		& $1.9\times 10^{11}$	& $1.7\times 10^{10}$		& $2.8\times 10^{9}$& $4.5\times 10^{9}	$	& 101		& 8.7		& 0.498					& 10		\\
\end{tabular}
\end{center}
\caption[Properties of the a set of galaxies with different ISM models at z = 0]
{Properties of the main halos at z =0.
The virial mass, gas mass within the virial radius ($R_{vir}$), cold gas mass, and stellar mass are calculated directly from the simulations.
The cold gas is defined as the sum of HI, $\Hmol$, and HeI.
In the primordial and metal simulations, the $\Hmol$ mass is zero and the hydrogen gas that would be molecular remains as HI.
$V_f$ was calculated for each galaxy by fitting an arctangent function to the circular velocities.
The optical radius of each of the galaxies is defined to be the radius at which the mean B-band surface brightness is 25 magnitudes per square arcsecond.
The SFR at z = 0 for each of the galaxies was calculated from the SUNRISE-generated H$_\alpha$ emission, using the same conversion as \citet{Kennicutt98}.
The ``\% Gas Lost'' refers the the mass of supernova-heated gas expelled beyond the virial radius divided by the total mass of gas particles ever accreted onto the disk of the galaxy (see \S 3.4 for details).
\vspace{0.5cm}

$^1$ Appears with primordial cooling and an additional low temperature cooling model in \citet{Governato10}.  The galaxies are labeled as DG1(h516) and  DG2 (h799).\\
$^2$ Appears with the Primordial ISM model in \citet{Brooks11}.  In \citet{Brooks11}, both h603 and h986 are at lower resolution.\\
$^3$ Appears with the Metal ISM model in \citet{Oh11b} as DG1 (DG2).\\
$^4$ Appears with the Metal ISM model in \citet{Pontzen11}.\\
$^5$ Appears with the $\Hmol$ ISM model in \citet{Christensen12}.\\
$^6$ Appears with the $\Hmol$ ISM model in \citet{Governato12}.\\
$^7$ Appears with the $\Hmol$ ISM model in \citet{Munshi12}.\\
$^8$ Appears with the Metal ISM model in \citet{Shen13a} as "Bashful".
}
\end{table*}

%** Halo selection
The central galaxies in each simulation were first identified using AHF \citep[AMIGA Halo Finder]{gill04, knollmann09}. 
AHF adopts overdensities as a function of redshift from \citet{gross97}, with an overdensity of $\sim$100 with respect to the critical density, $\rho_{crit}$, at $z=$0. 
The environment density at z$=$0 is typical of field halos with moderate over-underdensity $\delta \rho$/$\rho$ = -0.3--0.3 measured over a sphere of radius 3 h$^{-1}$ Mpc. 
The virial mass, gas mass, cold gas mass (defined as the combined mass of HI, HeI and $\Hmol$, when appropriate), and stellar mass of the central galaxies (defined as the stellar mass within the virial radius that is not within satellites) are listed in columns 1-4 of Table 1.
The galaxies simulated with the $\Hmol$ ISM model were previously analyzed in \citet{Munshi12}, in which they were shown to lie along the observed stellar mass-halo mass relation.
The asymptotic circular velocities of the galaxies are listed column 5 of Table 1 and the method used to calculate them is described in \S\ref{sec:globalProperties}.

%Observed Properties
In order to properly compare the outputs from our simulations to real galaxies, we calculated the star formation rate (SFR) indicators and the gas surface densities of the galaxies in an observationally motivated fashion (Table 1).
%Sunrise
We used the Monte Carlo radiation transfer code, {\sc Sunrise} \citep[v 3.6][]{Jonsson06} to generate artificial photometric images and spectral energy distributions (SEDs) of the simulated galaxies.
This radiative transfer calculation includes dust scattering as part of the ray-tracing analysis.
We assume a dust-to-gas ratio of 0.4 to calculate the dust mass for each of the SPH particles.
Figure~\ref{fig:morphology} show the four galaxies simulated with the $\Hmol$ ISM model at a redshift of zero.
The initial conditions for the two dwarf galaxies (h799 and h516) form disks with irregular distributions of recent star formation.
The initial conditions for the two larger galaxies (h986 and h603) form spirals and in once case (h986 with the metal ISM model; not shown here) a strong bar.

% ** SFR Indicators
{\sc Sunrise} enabled us to measure spatially-resolved SFRs of the galaxies from the 24$\mu$m and FUV emission observational indicators.
We converted these indicators into SFRs using the formula listed in \citet{Leroy08} in order to make an accurate comparison to the resolved Kennicutt-Schmidt relationship generated for a large set of nearby galaxies in \citet{Bigiel08} (see \S\ref{sec:ISM}).
In order enable direct comparisons with a larger sample of observed galaxies, we used H$\alpha$ emission to determine the SFR over the entire disk of the galaxy (column 7 of Table 1).
The B-band optical radii of the each of the galaxies are listed in column 6 of Table 1.

%** Gas Surface Densities
We measured the gas surface densities of the main halos in a way comparable with THINGS \citep[The HI Nearby Galaxies Survey,][]{Walter08} observations through the use of simulated HI velocity cubes.
We created a velocity cube for each galaxy by rotating it to $45^{\circ}$ angle and convolving each gas particle with the smoothing kernel and HI fraction calculated in {\sc Gasoline}.
In order to mimic typical THINGS resolution and sensitivity, we proceeded as if the galaxies were at a distance of 5 Mpc.
We used 1.5'' pixels and applied beam smearing with a full width at half max of 10'' to the data.
We divided the data into 128 velocity channels, which had widths of 2.6 km/s for the two spiral galaxies and widths of 1.3 km/s for the dwarf galaxies.
Finally, we  made a sensitivity cut and ignored all emission from cells below the 2$\sigma$ noise limit for the THINGS observations, in which $\sigma = 0.65$ milli-Jansky/beam.
This velocity cube was used to create images of the HI surface density and to compare to the resolved Kennicutt-Schmidt relation (\S\ref{sec:ISM}).
Figure~\ref{fig:pretty_pic} shows the mock-observations of both the ISM and stellar light for one of the spiral galaxies (h986) simulations with the $\Hmol$ ISM model at a redshift of zero.
 
 %%%%%%%%%%%%%%%%%%%%%%
\section{Results}
\subsection{Mass Distribution at z$=$0}\label{sec:globalProperties}
Previous simulations \citep{Governato10, Guedes11, Pilkington11, Stinson13} have suggested that changes to the ISM, in particular the star forming gas, can change the efficiency of feedback and, consequentially, the mass profile of galaxies.
Here, we systematically examine how the different models of the ISM and their interaction with the SN feedback can change the mass distribution in galaxies, as seen in their density profiles and rotation curves.

%--------------------------------------- Circular Velocities
\begin{figure*}
\begin{center}
$
\begin{array}{cc}
\includegraphics[width = 0.5\textwidth]{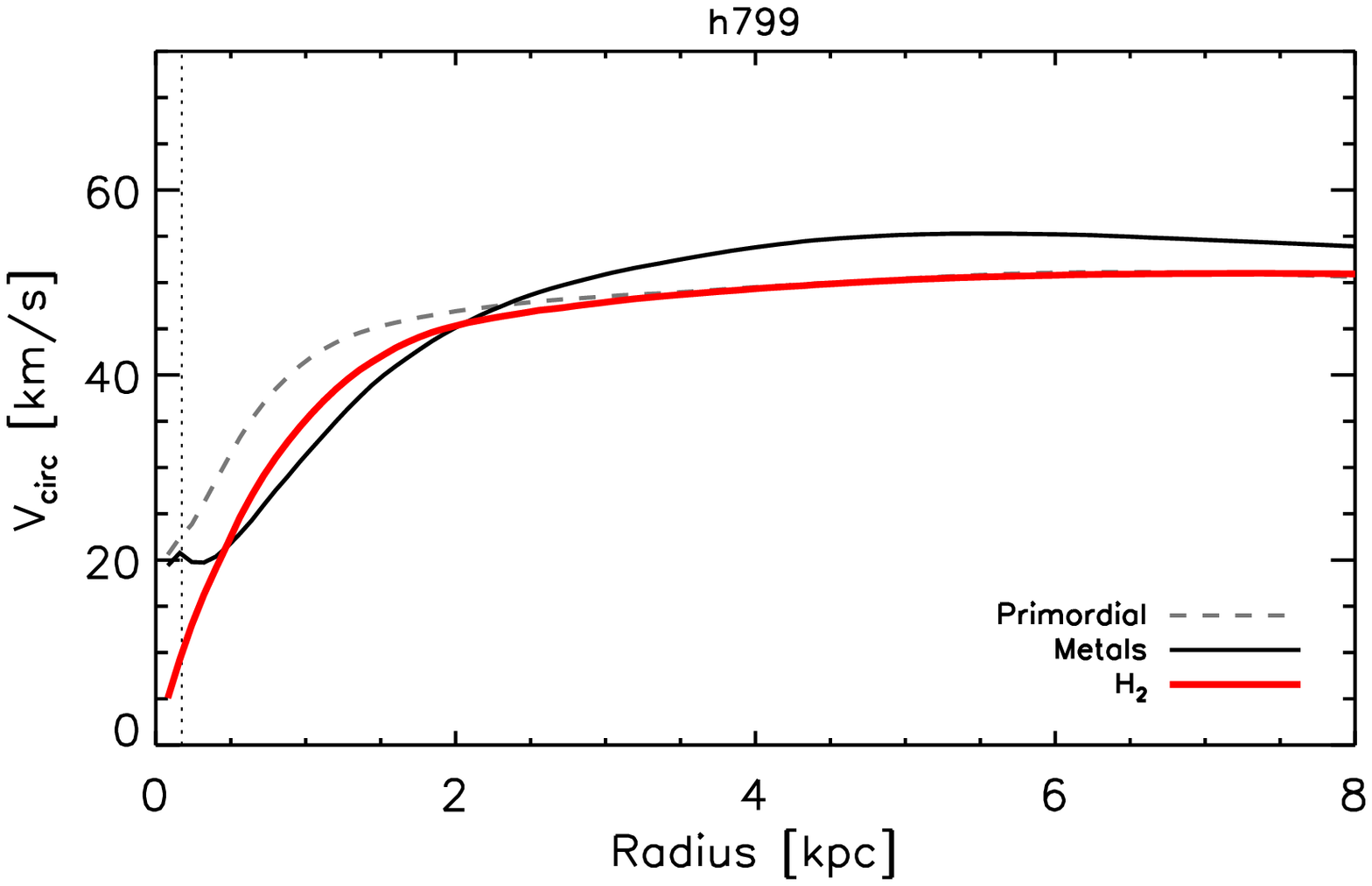}&
\includegraphics[width = 0.5\textwidth]{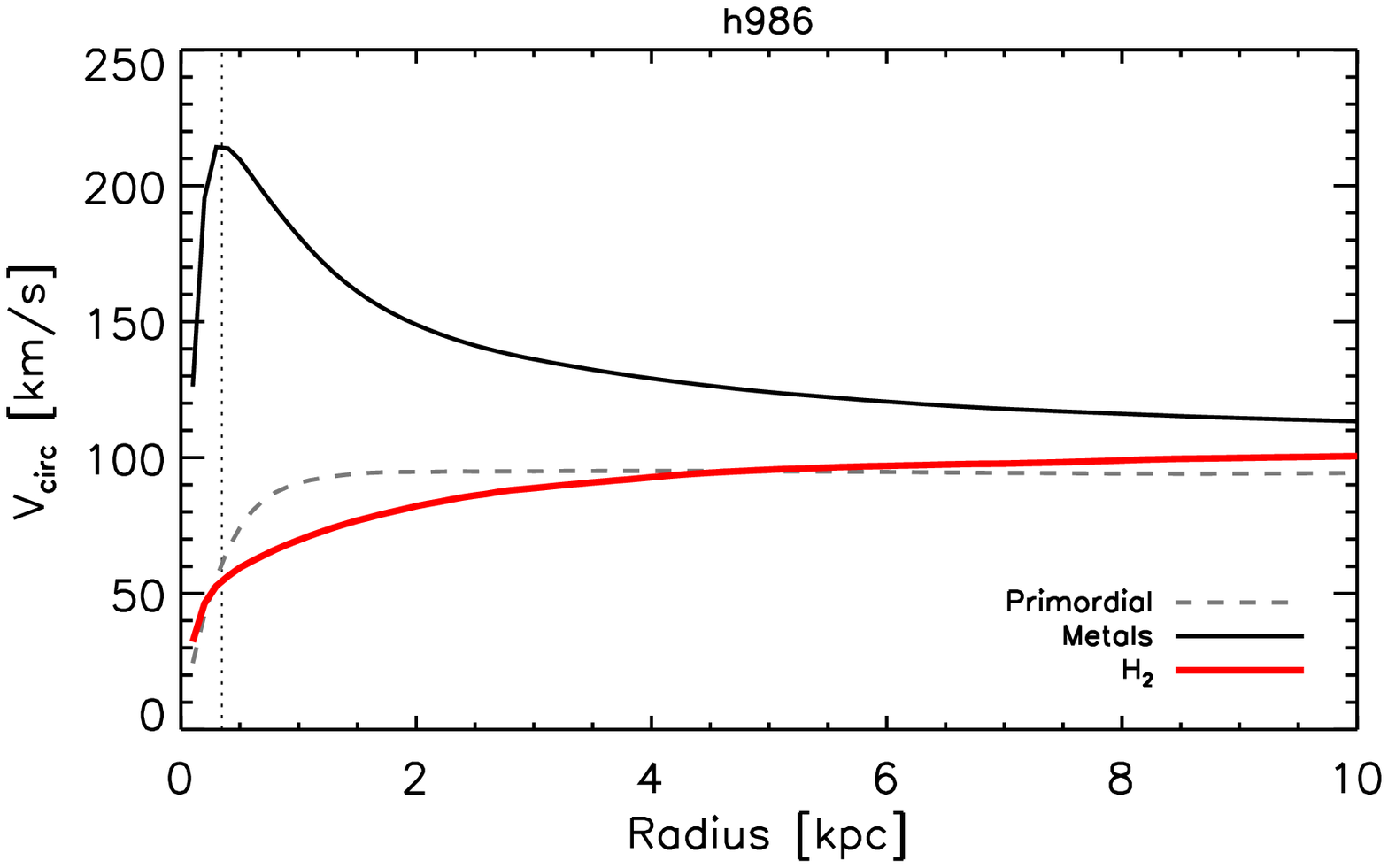}\\ 
\includegraphics[width = 0.5\textwidth]{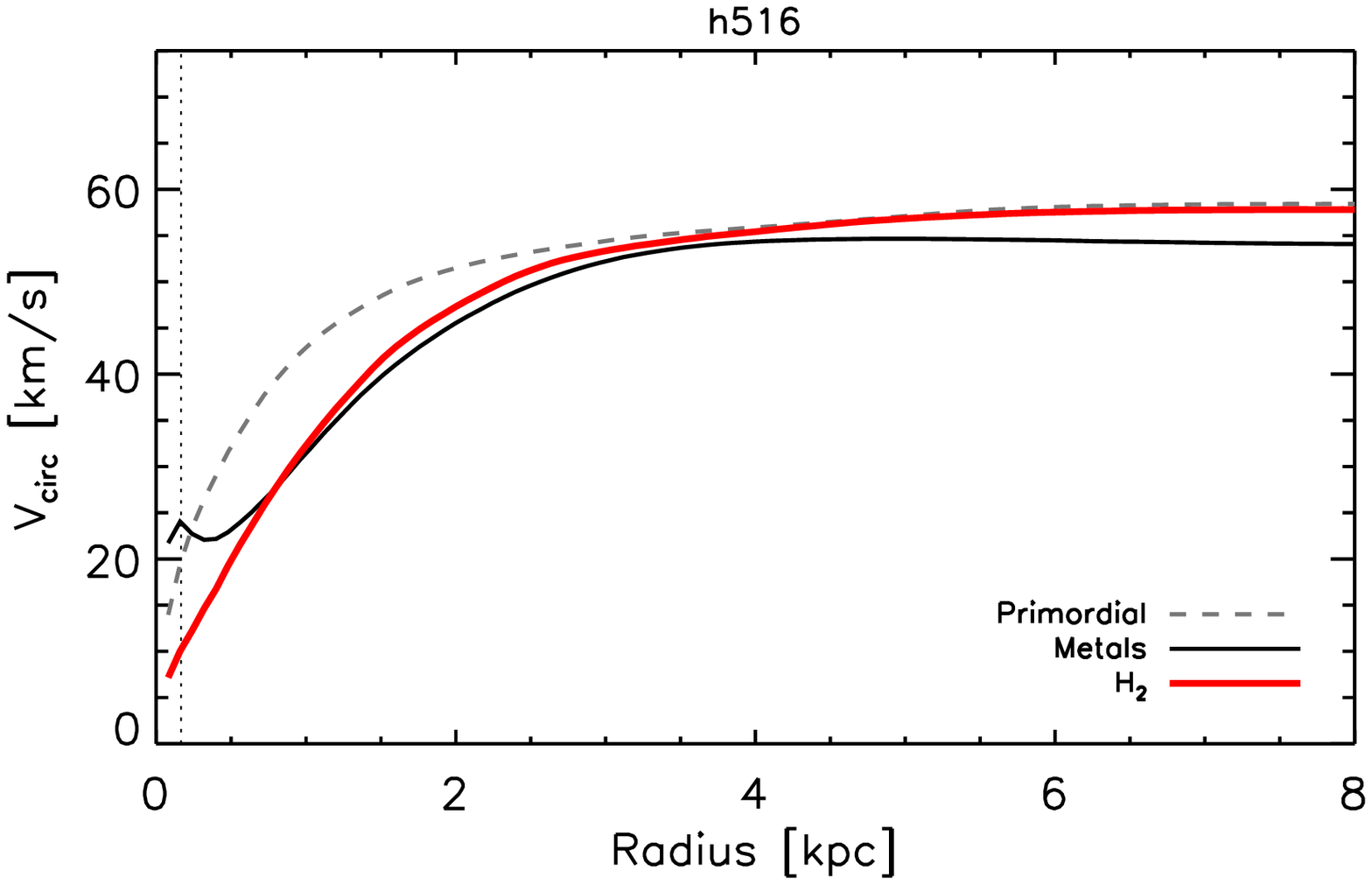}&
\includegraphics[width = 0.5\textwidth]{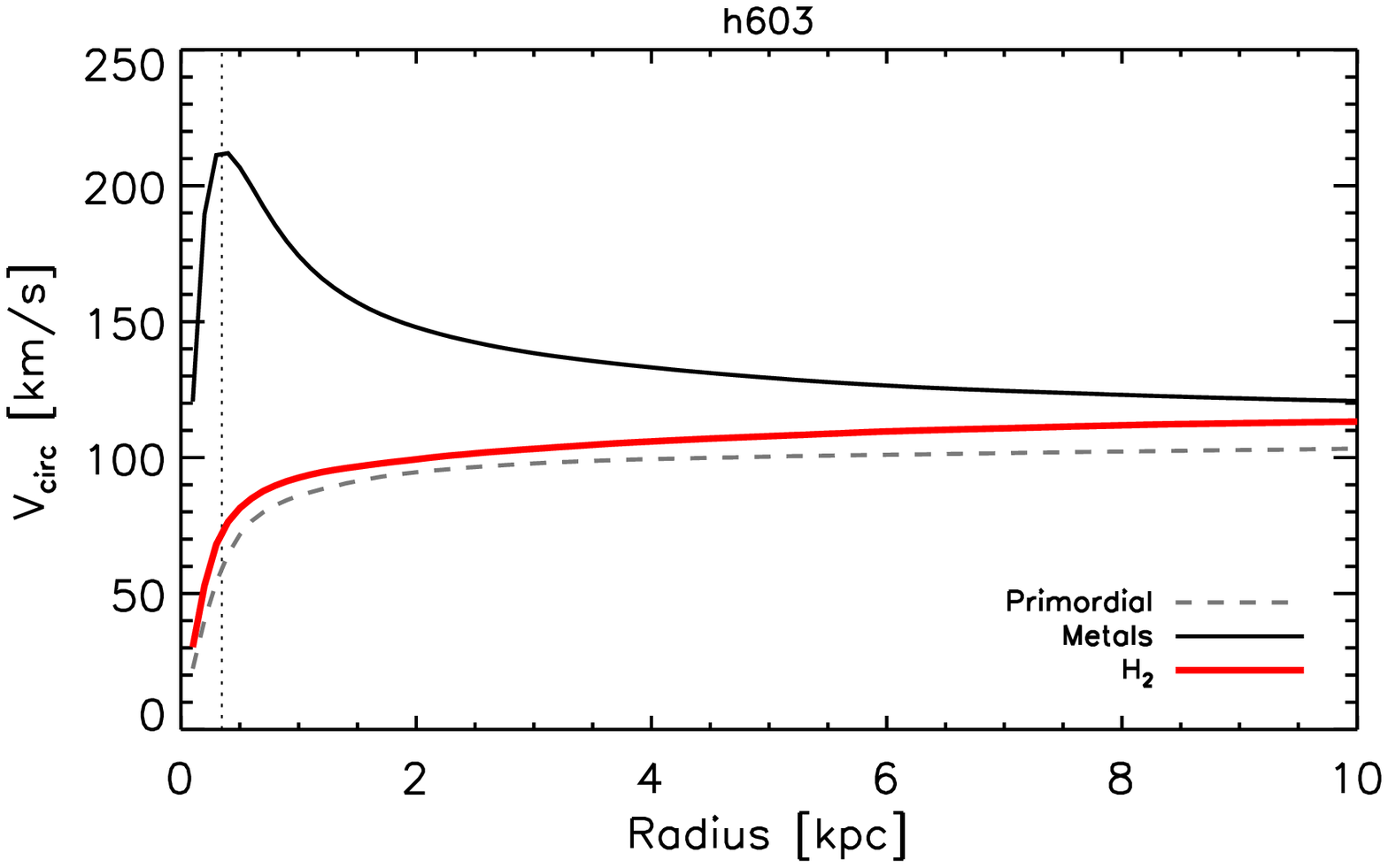}
\end{array}$
\end{center}
\caption[Circular velocities for a set of galaxies simulated with different ISM models]
{
The circular velocity profiles of the galaxies.
The left panels show the dwarf galaxies and the right panels show spiral galaxies.
Red lines represent simulations with the H$_2$ ISM model; black lines represent simulations with the metal ISM model; grey dashed lines represent simulations with the primordial ISM model.  
The vertical dotted line indicates the radius at twice the softening length.
The circular velocity curves of the dwarf galaxies have qualitatively the same shape across the different ISM models.
In the dwarf galaxies simulated with the metal ISM model, the flattening of the circular velocity curve at small radii is the result of a large stellar cluster at the center of the galaxy.
Simulations of the spiral galaxies with the $\Hmol$ or primordial ISM models have rising circular velocity curves, in contrast to the steeply peaked circular velocity curves of the simulations with the metal ISM model.
 }
\label{fig:rotcurve}
\end{figure*}

Figure~\ref{fig:rotcurve} shows the circular velocity curves of the four main halos simulated using the different ISM models.
These curves show the circular velocity at each radii calculated from the total mass contained within that radii. 
The circular velocity curves of the spiral galaxies show strong variation with the different ISM models.
Only the primordial and $\Hmol$ models produced continually rising rotation curves.
The two spiral galaxies simulated with the metal ISM model have a central peak.
A comparison sample of observed galaxies is available in \citet{deBlok08}, which provides a set of HI-determined rotation curves for galaxies from THINGS.
The gentle rise of the circular velocity curves for the galaxies simulated with either the primordial or $\Hmol$ model is comparable to rotation curves of the observed galaxies.
The central peak of the galaxies with the metal model, however, indicates a much higher central concentration and a build-up of low-angular momentum baryons in conflict with observed constraints.

In contrast, we found that all of the ISM models produced continually rising circular velocity curves for the dwarf galaxies.
The circular velocity curve of the dwarf galaxies produced with the metal ISM model flatten near the center because of the presence of a large stellar cluster.
This cluster is formed in one merger-driven burst further out in the disk;  it does not follow the classic bulge-formation paradigm.
The similarity of the rotation curves for the dwarf galaxies, aside from the cluster, is expected.
The low metallicity and $\Hmol$ content of these galaxies would result in relatively small differences between the ISMs produced by the three models. 

%---------------------------------------- DM profiles
\begin{figure*}
\begin{center}
$
\begin{array}{cccc}
\includegraphics[width = 0.5\textwidth]{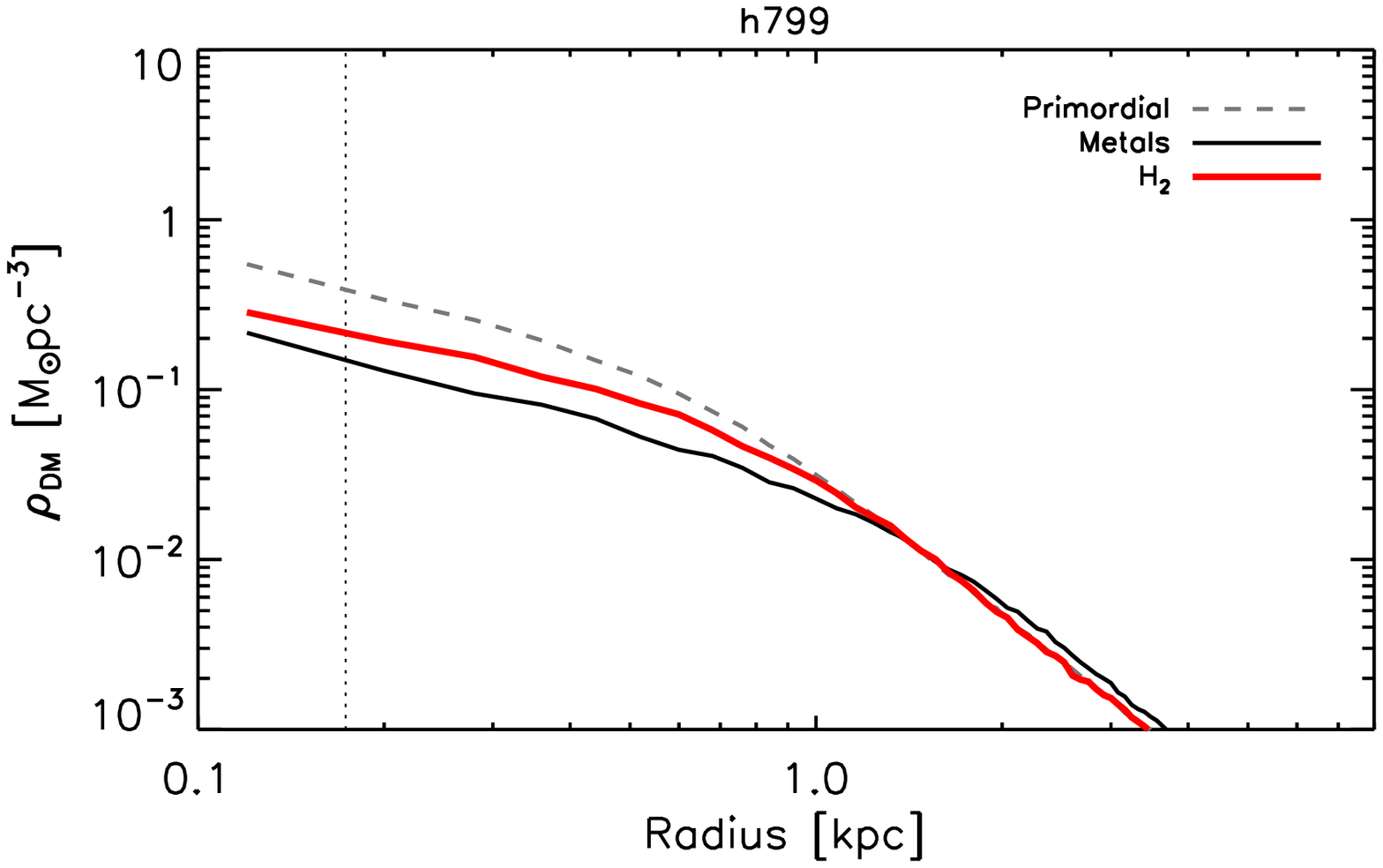}&
\includegraphics[width = 0.5\textwidth]{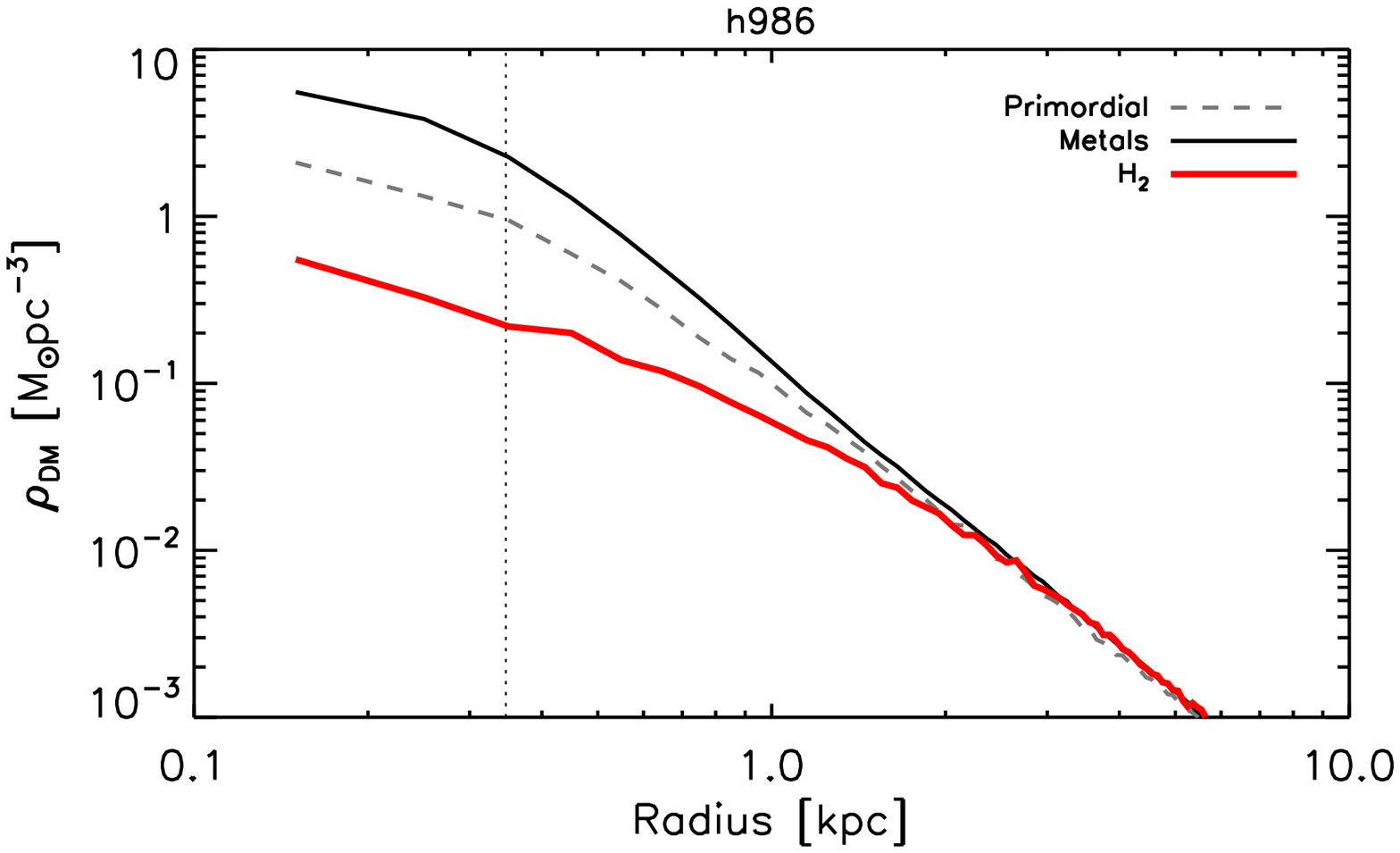}
\end{array}$
\end{center}
\caption[Dark matter profiles for a set of galaxies simulated with different ISM models]
{
Dark matter density profiles of the galaxies.
The left panel shows a representative dwarf galaxy (h799) and the right panel shows a representative spiral galaxy (h986).
Red lines represent simulations with the H$_2$ ISM model; black lines represent simulations with the metal ISM model; grey dashed lines represent simulations with the primordial ISM model.  
The vertical dotted line indicates the radius at twice the softening length.
The central densities of the dwarf galaxies are similar across all ISM models.
The central concentration of the spiral galaxies is lowest for the $\Hmol$ ISM model, followed by the primordial ISM model, and highest for the metal ISM model.
This pattern is similar to the changes to the circular velocities with ISM model.
 }
\label{fig:DMprof}
\end{figure*}

In addition to the rotation curves, the changes to the mass distribution are also apparent in the DM profiles (Figure~\ref{fig:DMprof}).
The DM profiles generated with different ISM models are shown for a representative dwarf and spiral galaxy.
Similarly to the rotation curves, the most dramatic changes to the DM profiles occurred in the spiral galaxies: the DM is most concentrated for the metal simulations and least in the $\Hmol$ simulations with the primordial simulations lying in-between.
The relatively low concentrations of the $\Hmol$ and primordial simulations compared to the metal simulation is consistent with the lack of peaked rotation curves.
The changes to the DM profile indicate that the reduced circular velocities seen in Figure~\ref{fig:rotcurve} are not only the result of reduced central baryonic mass but also to an alteration in the galaxies's DM distributions.
The dwarf galaxies, however, experienced only slight changes to their DM profiles and, in contrast to the spiral galaxies, the metal ISM model produced the smallest central concentrations in both h799 and h516.
As with the rotation curves, the similarities between the DM profiles can be explained by the low metallicities and molecular content of these galaxies, which reduces the differences between the ISM models.

%------------------------------------- Angular Momentum 
\begin{figure*}
\begin{center}
 $
\begin{array}{cccc}
\includegraphics[width = 0.5\textwidth]{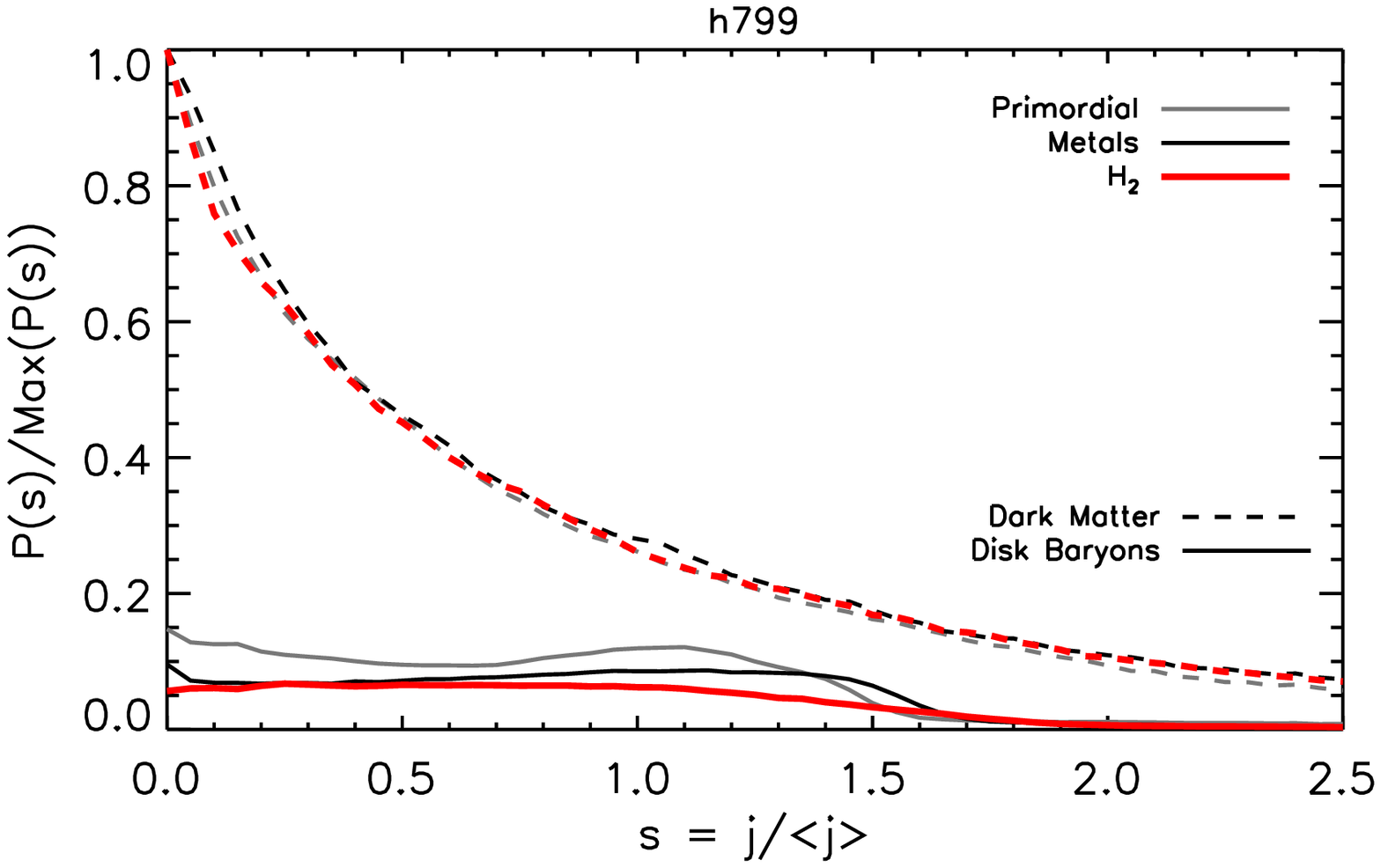} &
\includegraphics[width = 0.5\textwidth]{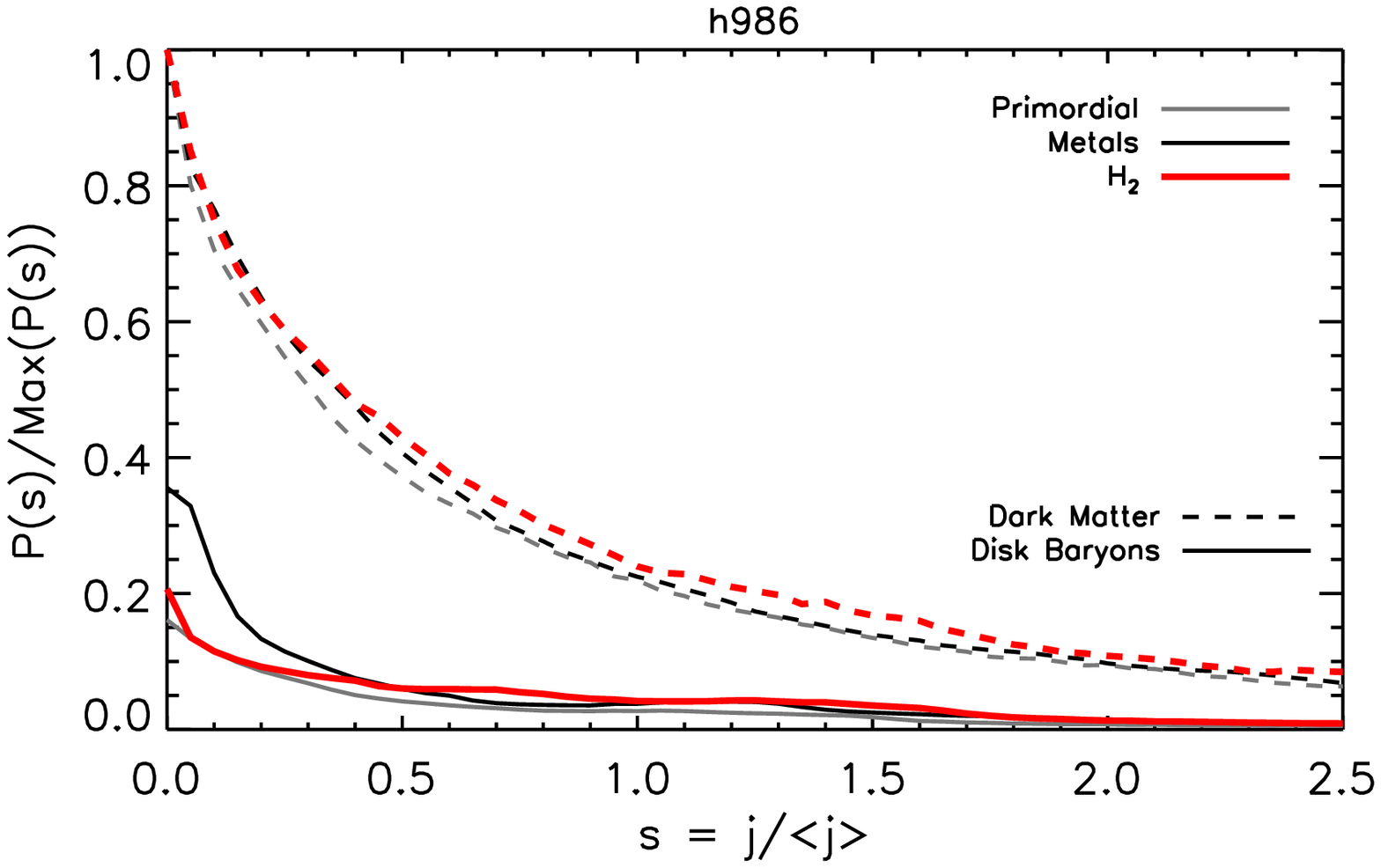}
\end{array}$
\end{center} 
\caption[Angular momentum distribution in a set of galaxies with different ISM models]
{
The AMD of the DM (dashed lines) and disk baryons (solid lines).  
Results for a representative dwarf galaxy (h799) appear in the left panel and results for a representative spiral galaxy (h986) appear in the right panel.
Red lines represent simulations with the H$_2$ ISM model; black lines represent simulations with the metal ISM model; grey lines represent simulations with the primordial ISM model.  
For both the dark matter and baryons, the distribution, P(s), is shown for the angular momentum, $j$, scaled by the mean angular momentum, $\langle j \rangle$.
These distributions of scaled angular momentum are then divided by the maximum of the distribution.
For the baryons, the P(s) is again divided by the cosmic baryon fraction to produce the final plot.
The baryons used for calculating the angular momentum lie within the optical radii and consist of the stars and cold gas (HI, HeI and $\Hmol$, when appropriate).
The changes to the AMD are consistent with the changes in the rotation curves.
In the spiral galaxies, the use of either the primordial or $\Hmol$ ISM model results in  decreased baryonic mass in the disk and a smaller mass of low-angular momentum baryons compared to the galaxies simulated with the metal ISM model.
In the galaxies with the metal ISM model, the AMD is slightly centrally peaked.
The angular momentum profiles of the dwarf galaxies are much more similar, although the galaxy simulated with the primordial ISM models has slightly more disk baryons and the galaxy simulated with the $\Hmol$ model has a slightly lower central concentration.
 }
\label{fig:jdist}
\end{figure*}

The changes to the rotation curves and DM profiles denote changes to the AMD.
Figure~\ref{fig:jdist} shows the AMD for both the DM and baryonic mass in a representative dwarf and spiral galaxy.
The baryons in the plot include all observable baryons, i.e., the stars and the HI, HeI, and $\Hmol$ gas (if applicable).  
The DM AMD was normalized by dividing by the maximum of the distribution, while the baryon AMD was divided by the maximum and then further scaled by the baryon fraction within the halo (so that a maximum value of 1.0 indicates that the halo has the cosmic baryon fraction).
The use of the metal ISM model resulted in a strong peak at zero angular momentum in the spiral galaxy.
This peak was caused by a relatively large amount of very low-angular momentum material.
In addition to lowering the central peak, both the primordial and $\Hmol$ ISM models resulted in less baryonic material in the disks of the spiral galaxies, indicating greater amount of baryonic mass lost from the simulations.
As was the case for the rotation curves and DM profiles, the AMDs of the dwarf galaxies show much less variation between models.
The angular momentum of simulated galaxies is known to be affected by stellar feedback in a variety of ways.
Stronger feedback can reduce angular momentum loss by delaying star formation such that mergers happen between more gas rich disks and by reducing the amount of star formation during the merger (much of which would be in a bulge component).
The angular momentum distribution can also be changed by feedback as outflows preferentially drive low-angular momentum material.
We examine the build-up of the stellar material in \S3.3 and the strength of feedback and the characteristics of the expelled gas in greater detail in \S3.4.

%%%%%%%%%%%%%%%%%%%%%%%%%%%%%%%
\subsection{Properties of the ISM}\label{sec:ISM}
In order to determine the causes of the differences between the central mass distributions resulting from the different models, we compared the properties of the ISM.
The gas surface densities for a spiral galaxy simulated with each of the ISM models are shown in Figure~\ref{fig:pretty_picISM}.
In this figure, the logarithmic HI surface densities at the THINGS resolution and sensitivity are shown in grey-scale.
For the $\Hmol$ model ISM, the logarithmic $\Hmol$ surface density is shown in the red contours.
This figure shows the variation in the gas distribution produced by the different ISM models.
With increased amounts of cooling, the gas distribution becomes both clumpier and more extended.

%------------------------------------ ISM Image --------------------
\begin{figure*}
\includegraphics[width=1.0\textwidth]{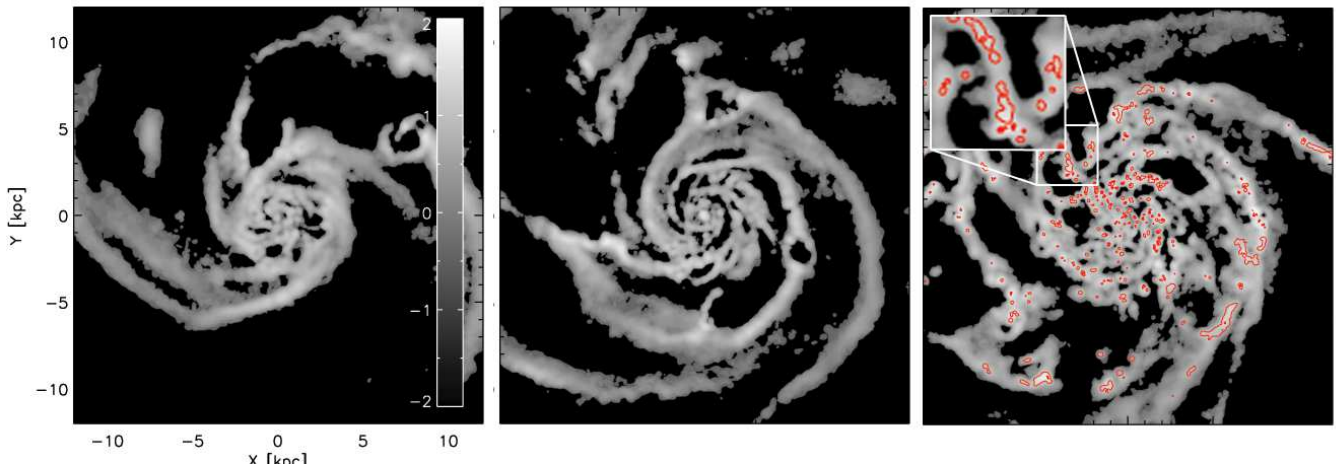} 
\caption[Gas surface density of an L$^*$ galaxy]
{Gas surface density of a spiral galaxy (h986) simulated with each of the ISM models: left (primordial), middle (metals), right ($\Hmol$).
The gray scale is the HI surface density at the THINGS resolution and sensitivity, ranging from log $\Sigma_{\mathrm{HI}}$ = -2 to 2 amu/cc.
In the right panel, the red contours show the $\Hmol$ surface density at log $\Sigma_{\Hmol}$ = -2, 0, and 2 amu/cc.  
The clumpiness, amount, and extent of disk gas increase from left to right.
}
\label{fig:pretty_picISM}
\end{figure*}

%-------------------------------- Gas surface density/density ---------------------------
\begin{figure}
\includegraphics[width=0.5\textwidth]{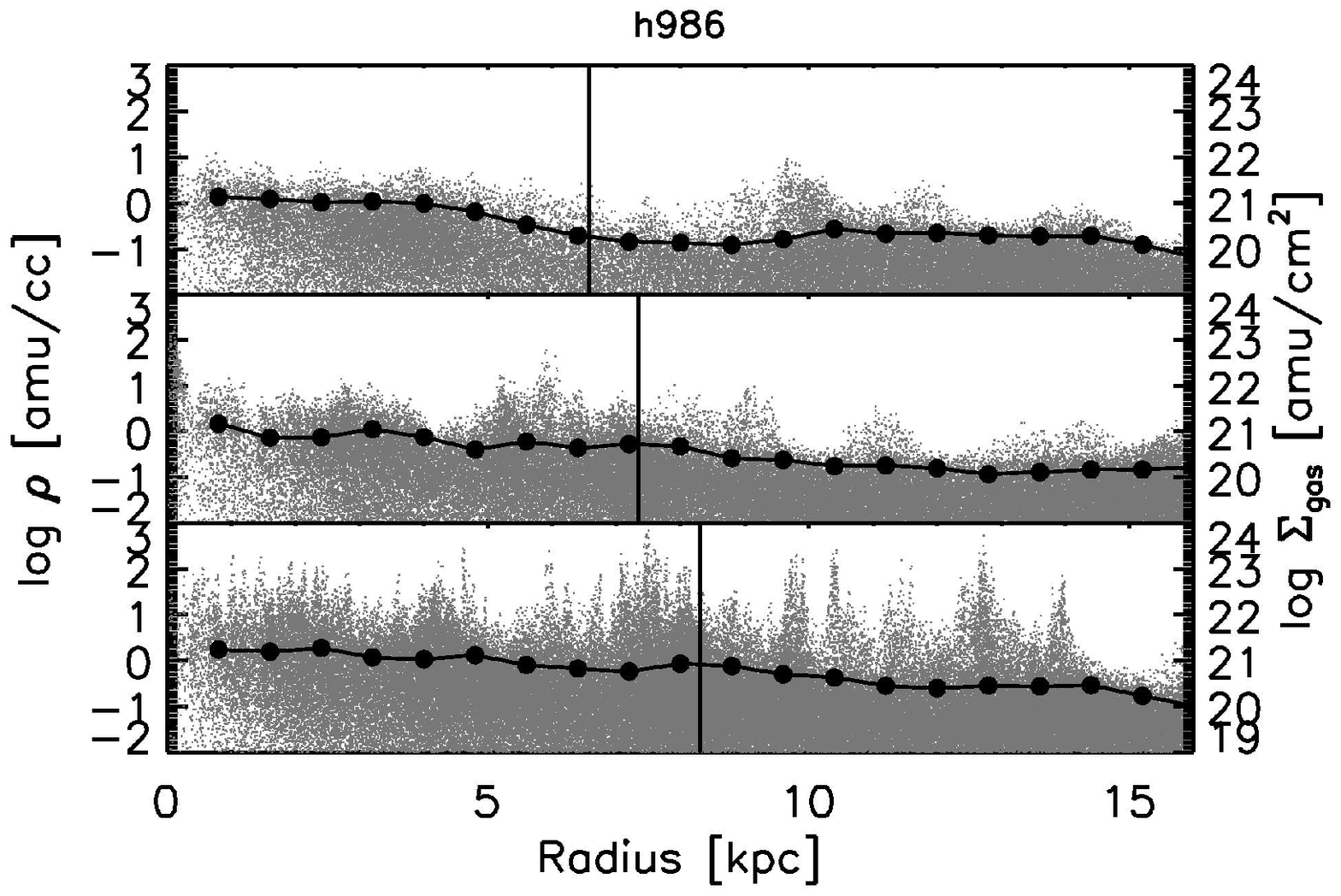}
\caption[Gas density and surface density as a function or radius]
{Density and azimuthally averaged surface density of gas as a function of radii for a representative spiral galaxy (h986) simulated with the different ISM models 
The top panel shows the primordial ISM model, the middle panel shows the metal ISM model, the bottom panel shows the $\Hmol$ model.
Grey points correspond to the left axis and represent the density of gas particles.
The black curves correspond to the right axis and represent the radially-binned average surface densities of the gas.
Vertical lines mark the optical radii.
The increased cooling from metals and the shielding in the $\Hmol$ model resulted in clumpier gas 
although the average surface densities were similar for the different simulations. 
Similarly, the metal ISM model shows more density peaks than in the primordial ISM model, although without the presence of shielded gas, the effect is less strong.
}
\label{fig:SD_dens}
\end{figure}

Changes to the ISM model affect not only the densities of the star forming gas but also the densities of nearby gas particles.
The effect of the different ISM models on the gas density and azimuthally-averaged surface density across the disk of the galaxy are shown in Figure~\ref{fig:SD_dens}.
In this plot, the three panels represent the different ISM models.
Within the optical radii there is little change in the average surface density across simulations.
However, the densities of individual particles vary substantially between the models.
The increased cooling in the metal line model and the shielding in the $\Hmol$ model resulted in the gas particles reaching progressively higher densities.
High-density star forming regions appear in the plot as spikes in the grey points where gas particles at similar radii have  especially high densities.
In both the primordial and metal ISM model, few of the gas particles have densities above the star formation threshold of 10 amu/cc. 
This limited amount of star forming gas is consistent with the low star formation rates for these galaxies at z= 0 (Table 1) and results from a combination of lower surface densities and the relatively high value of $c^*$, which ensures that star formation proceeds quickly once the gas has passed the density threshold.  
At earlier times, a much greater fraction of the gas was star forming.

The formation of these clumps allows for star formation to continue throughout a greater extent of the disk.
There is a slight increase in the optical radii (marked by the vertical lines and listed in Table 1) in the ISM models with more cold gas.
This trend in greater optical radii at z = 0, indicates a greater spatial extent of on-going star formation.
As shown in \citet{Christensen12}, the additional cold gas allows star forming regions to form farther out in the disk of the galaxy in the $\Hmol$ simulation.
This trend of increased clumpiness as first metal-cooling and then shielding (incorporated with the $\Hmol$) are added also appears in the dwarf galaxies, although the low-metallicity of these galaxies means that the difference between the metal and primordial models is much smaller.

%------------------------------------------- Power Spectrum ----------------------------
\begin{figure}
\begin{center}
\includegraphics[width = 0.5\textwidth]{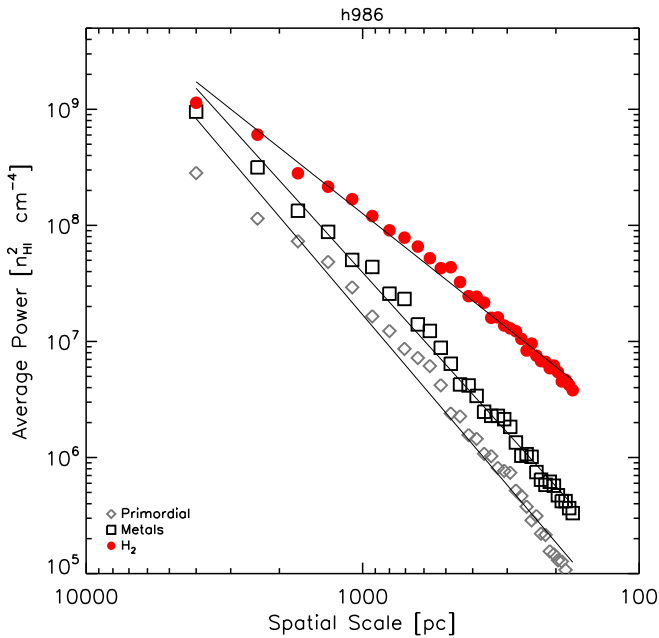}
\end{center}
\caption[Power Spectrum of the ISM for a set of galaxies simulated with different ISM models]
{
Power spectrum of the HI surface density in a representative spiral galaxy (h986).
Filled red circles represent simulations with the H$_2$ ISM model; black empty triangles represent simulations with the metal ISM model; grey empty diamonds represent simulations with the primordial ISM model.  
The 2-dimensional Fourier transform was calculated on the face-on HI surface density and a power-law was fit to the power spectrum (solid lines).
The introduction of $\Hmol$ produces a shallower slope than the other two ISM models, indicating that the gas has more power at small scales (i.e. clumpier).
}
\label{fig:powerspectrum}
\end{figure}

We quantified the clumpiness of the gas for each of the ISM models using the power spectrum of the HI surface density distribution (Figure~\ref{fig:powerspectrum}).
We calculated the power spectrum for the each of the main halos with the different ISM models by computing a 2-dimensional Fourier transform of the face-on HI surface density using a Fast Fourier Transform algorithm. 
For the HI surface density, we used the first moment of the HI	 velocity cube.
However, we did not apply the THINGS beam smoothing or make a sensitivity cut for this analysis as most observations of the power spectra of external galaxies are done with resolutions comparable to or higher than our softening length.
We truncated the power spectrum at the force softening length (170 pc in the spiral galaxies).
In all simulations, the power spectra can be fit with power laws.
The power indicates the relative amount of energy at different scales with shallower slopes indicating greater the energy on small scales, i.e. greater clumpiness.
The exponents of the power law fit for h986 (h603) are $-2.8$ ($-3$) in the primordial cooling model,  $-2.6$ ($-2.7$) in the metal cooling model, and $-1.9$ ($-2.1$) in the $\Hmol$ model.
The fits for observed galaxies on scales greater than 200 pc are consistent with these values and have range between $-1.5$ and $-3$ \citep{Stanimirovic99, Elmegreen01, Begum06, Dutta08,  Dutta09}.
Therefore, we cannot use the power spectrum to distinguish between the accuracy of these models.
However, it is clear that the shielding included in the $\Hmol$ model produced a clumpier gas with more power at small-scales.
We see no strong differences between the power spectra of galaxies with the primordial or metal ISM models.
The dwarf galaxies show similar trends in their power spectra, although the differences are smaller because of the lower metallicities.
\citet{Pilkington11}, in contrast, detected substantially more power at large scales in a dwarf galaxy run with metal line cooling compared to a dwarf galaxy without it.
That galaxy, however, was caught in a state without any on-going star formation, which may account for the enhanced formation of larger-scale structure, such as spiral arms.

%%%%%%%%%%%%%%%%%%%%%%%%%%%%
\subsection{Star Formation}\label{sec:decomp}
The differences in the ISM discussed in the previous section lead to changes in the star formation.
In addition to having lower central concentrations of baryons, the spiral galaxies with the primordial or $\Hmol$ ISM model have about half the total stellar masses than the galaxy produced with the metal ISM model (Table 1).
It is also apparent from images of these galaxies (Figure~\ref{fig:morphology}) that the spiral galaxies with the primordial or $\Hmol$ ISM models have smaller central stellar masses, i.e. smaller bulges.
Finally, as discussed in the previous section, greater amounts of cooling or shielding in the ISM models produced denser clumps where star formation could take place and, consequentially, more extended star formation.
In order to examine how the different models of the ISM affect the star formation in galaxies, we compared the star formation histories and star formation laws of the simulations.

%------------------------------------- Star Formation History
\begin{figure}
\begin{center}
\includegraphics[width = 0.5\textwidth]{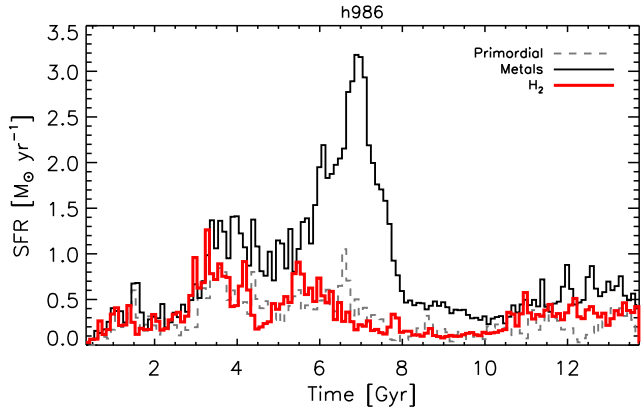}
\end{center}
\caption[Star Formation History]
{
Star formation histories of the representative spiral galaxy, h986, produced by each of the models.
The red line represents the simulation with the H$_2$ ISM model; the black line represents the simulation with the metal ISM model; the grey dashed line represents the simulation with the primordial ISM model.  
There is a strong decrease in the star formation between 6 and 8 Gyrs for the primordial and $\Hmol$ ISM models compared to the metal ISM model.
The different models do not, however, have a strong impact on the amount of stars formed before 3 Gyrs.
}
\label{fig:sfh}
\end{figure}

Figure~\ref{fig:sfh} shows the star formation history for one of the spiral galaxies.
We find that both the $\Hmol$ and primordial ISM models reduce the total stellar mass of the spiral galaxies, as can also be seen by comparing the data in Table 1.
In both galaxies, the star formation in these two models is especially reduced compared to the metal ISM model between 6 and 8 Gyrs.
This decrease is likely the result of greater gas-loss from feedback in the primordial or $\Hmol$ ISM models.
The reduced gas mass in the galaxy limits the amount of star formation in both of these models.
The amount of gas loss is further discussed in \S\ref{sec:snfb}.
The comparatively higher rates of star formation between 6 and 8 Gyrs in the metal ISM model directly affects the star formation at the center of these galaxies.
During this time, 49\% of the stellar mass within the central kpc was formed in the metal ISM model, compared to 28\% and 23\% for the primordial and $\Hmol$ ISM models, respectively.

The use of a $\Hmol$-dependent star formation law has been put forth as a way to reduce early star formation by raising the effective density threshold for star formation in low-metallicity gas \citep[e.g.][]{RobertsonKravtsov08,Gnedin10,Krumholz11a}.
However, we do not see a significant reduction of the stellar mass prior to 3 Gyrs in the simulations with the $\Hmol$ ISM model.
After this point in time, the metallicities of the galaxies have increased enough to reduce its effect on the star formation law.
As discussed in \citet{Christensen12}, star formation in our simulations is strongly regulated by feedback so the reduced star formation efficiency in low metallicity gas has a relatively small effect.

%------------------------------------- K-S law ------------------------
\begin{figure*}
\includegraphics[width=1.0\textwidth]{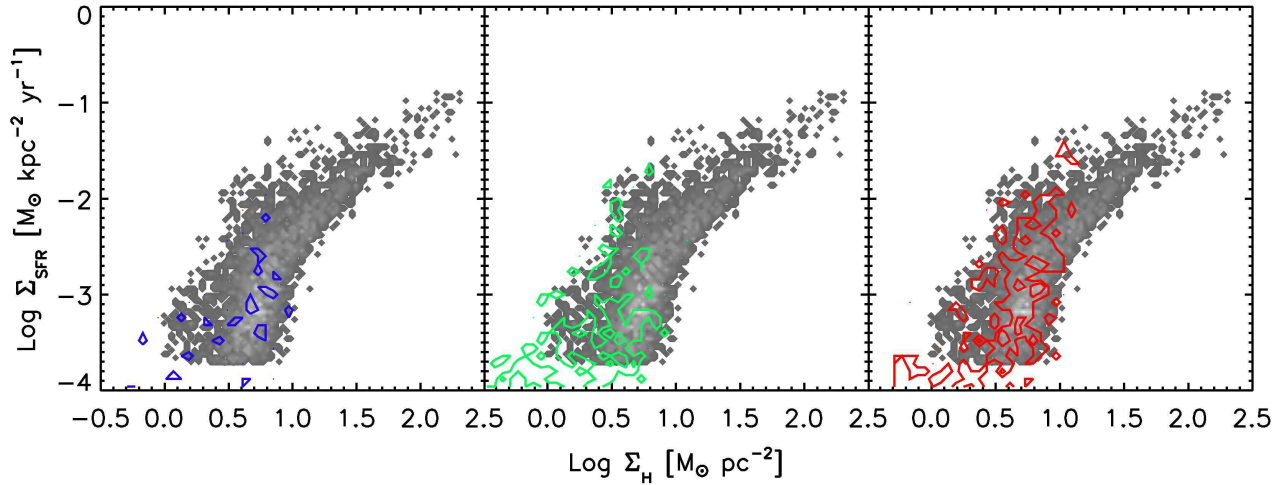}
\caption[Resolved K-S relation for a set of galaxies with different ISM models]
{Resolved K-S relation for the three different models of the ISM: left (primordial), middle (metals), right ($\Hmol$).
These plots show the SFRs and total gas (both HI and $\Hmol$) surface densities averaged over $750\pc \times 750\pc$ bins.
The grey-scale filled contours are observational data from \citet{Bigiel10} for local dwarf and spiral galaxies.
The colored contours represent the combined data for the simulated spiral and dwarf galaxies.
We calculated the SFRs from the SUNRISE-generated FUV and 24$\mu$m emission, the HI emission from the face-on velocity cube of average THINGS resolution and sensitivity, and the $\Hmol$ directly from the simulation.
All the ISM models produced a star formation law on kpc scales that is consistent with the observed data, although the metals ISM model may have a slightly less accurate fit.
}
\label{fig:sk}
\end{figure*}

Next we analyzed the star formation environments in each of the simulations.
In \S\ref{sec:ISM}, we established that higher cooling rates resulted in a clumpier ISM with the metal cooling ISM model than the primordial ISM model.
The addition of shielding in the $\Hmol$ ISM model results in even greater clumpiness than either of the two other models.
Here, we examine differences specifically in the star forming gas in order to relate them to differences in the star formation histories.
First, we computed the resolved K-S relation for the four galaxies produced with the three different ISM models, and compared them to the THINGS data outlined in \citet{Bigiel10} (Figure~\ref{fig:sk}).
In finding the resolved K-S relation for our sample of galaxies, we closely followed the observational methods.
The SFRs were calculated based on the 24$\mu$m and FUV emission from SUNRISE simulated observations.
The HI surface densities were taken from the zeroth moment of the HI velocity cube, which was computed at the same resolution and sensitivity of a standard THINGS observation.
The $\Hmol$ surface densities were found directly from the simulation output (i.e. we did not calculate it from the CO emission as was done in observed sample).
For each of the galaxies, we divided the SFRs and total hydrogen surface densities within the optical radius into $750\pc \times 750\pc$ bins.
The results are shown in Figure~\ref{fig:sk}.
The SFRs and total hydrogen surface densities of the spiral and dwarf galaxies with each of the different ISM models are represented by the colored contours.
The grey-scale filled contour map represents the observed data from \citet{Bigiel10}.
In general, the simulated data follow the same trend as the observed data.
Because of their low-mass, both the dwarf and low-mass spiral galaxies have surface densities lower than $\Sigma_{\textrm{H}} \sim 10 \Msun \textrm{pc}^{-2}$ and lie on the HI-dominated side of the K-S relation.
The metal ISM model may have a slightly steeper slope than either the observed galaxies or the two other simulations.
The difference, however, is slight and unlikely to have caused the differences in central stellar mass.
We, therefore, conclude that differences in the K-S relation between the simulations are not responsible for changes to the bulge.
The similarity between the results for the different ISM models also shows that the K-S relation is not a strong discriminator between our ISM models or star formation recipes in this regime.
While the ISM differ in their small scale structure, averaging over $750\pc \times 750\pc$ bins smooths over the differences between the models.

%------------------------------------------ Star Forming Gas -------------------------------
\begin{figure}
\begin{center}
$
\begin{array}{cc}
\includegraphics[width = 0.5\textwidth]{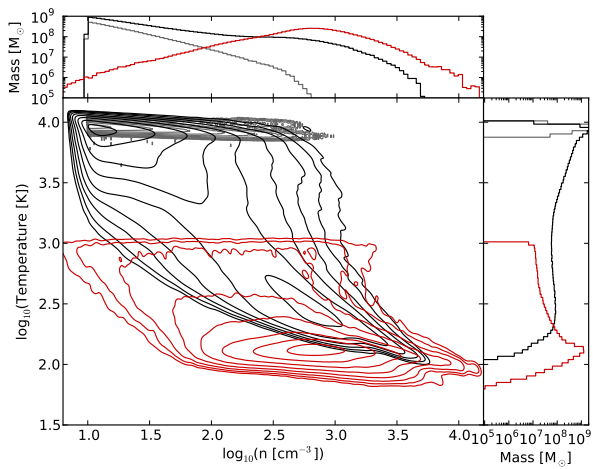}&
\end{array}$
\end{center}
\caption[Star forming gas]
{
Characteristics of the star forming gas in a representative spiral galaxy (h986) simulated with each of the different ISM models and star formation recipes.
The contours in the central panel show the distribution of gas particles that formed stars in density and temperature; they represent the mass of star particles formed and are spaced logarithmically.
The plots to the top and right show the histograms of the density and temperatures of the star forming gas.
The the red contours are for the $\Hmol$ ISM model, the black contours are for the metal ISM model, and grey contours are for the primordial ISM model.
Compared to the other two models, star formation in the $\Hmol$ ISM model takes place in colder gas because of the addition of dust shielding of HI and $\Hmol$ and denser gas because of the requirement for gas to be molecular to form stars. 
The ability of the gas in this model to become molecular and to cool to low temperatures at lower densities, mean that there is a comparatively large fraction of low-pressure star forming gas. 
}
\label{fig:sfg}
\end{figure}

In order to analyze the small scale differences, we next compared the densities and temperatures of the particles that form stars in the each of the models.
Figure~\ref{fig:sfg} shows the phase diagram for the gas particles immediately prior to forming stars.
This comparison illustrates how the star forming gas may be similar when averaged over $\sim 1$ kpc scales (as in the resolved K-S law shown in Figure~\ref{fig:sk}) but very different on the smaller scales.
Without any sources of low-temperature cooling, gas in the primordial ISM model formed stars at $10^4$K.
In the metal ISM model, the addition of metal line cooling resulted in star formation taking place at temperatures lower than $10^4$K.
However, gas in this simulation generally formed stars soon after crossing the 10 amu cm$^{-3}$ density threshold without cooling significantly first.
The addition of shielding of both HI and $\Hmol$ to the simulation with the $\Hmol$ ISM model resulted in a build-up of gas at $\approx100$ K.
Furthermore, the $\Hmol$-dependent star formation recipe caused star formation to take place in denser gas with higher $\Hmol$-to-HI fractions.
These densities are frequently higher than the current density where gas transitions from HI to $\Hmol$ because of the amount of high-redshift star formation that took place in lower metallicity gas and because the delay in star formation after $\Hmol$ has formed allows for further collapse.
In this model, there is also relatively more low-pressure star forming gas because the gas is able to become molecular and reach cold temperatures at lower densities than in the metal ISM model.
The presence of cold, dense, but relatively lower pressure star forming regions enables the formation of a clumpier ISM and can affect the efficiency of supernova feedback, as further discussed in the following section.

%%%%%%%%%%%%%%%%%%%%%%%%%%%%%%%%%%%%%%%%%%%%%%%%%%
\subsection{Feedback-Induced Gas Loss}\label{sec:snfb}

These changes to the ISM and, especially, the star forming gas have the potential to affect the distribution of gas throughout the galaxy by changing the efficiency of gas loss from feedback.
In order to study the efficiency of feedback, we tracked gas particles over the history of the simulation.
To accomplish this, we followed the history of the most massive progenitor galaxy and traced when gas particles were accreted to it, formed stars, and were ejected. 
In order to identify the gas particles ejected from the galaxy by SNe, we first identified the gas particles that had been part of the disk of the galaxy ($n \geq 0.1$ amu cm$^{-3}$, T $\leq 2 \times 10^4$ K and with a vertical distance to the central plane of the galaxy of less than 10 kpc) at any output.
We then looked for subsequent outputs in which the gas particle 1) was no longer part of the disk and 2) had been heated by a SN since the time it was last identified as being in the disk.
These two criteria defined the gas particles {\it ejected} from the disk.
We also defined a subset of the ejected gas to be the gas {\it expelled} from the main halo by selecting particles that traveled beyond the virial radius after being heated by SNe.
This gas will only ever again become a part of the galaxy is the virial mass is substantially increased following the expulsion.  

This method of identifying outflows is limited by the timestep resolution.
The outputs for our simulations are spaced at approximately 350 Myr intervals.
Therefore, this method does not identify particles that were ejected from the disk by SNe and then re-accreted onto the disk between two timesteps.
Since the simulations all share similar time resolution, however, the results may be used to compare them, even if the accuracy of the absolute amounts of ejected gas is limited.
The total fraction of gas ever accreted onto the disk which is later expelled from each of the four galaxies is listed in Table 1 in column 8.

%------------------------------------- Cumulative Gas Loss ----------------------------
\begin{figure*}
\begin{center}
$
\begin{array}{cc}
\includegraphics[width = 0.5\textwidth]{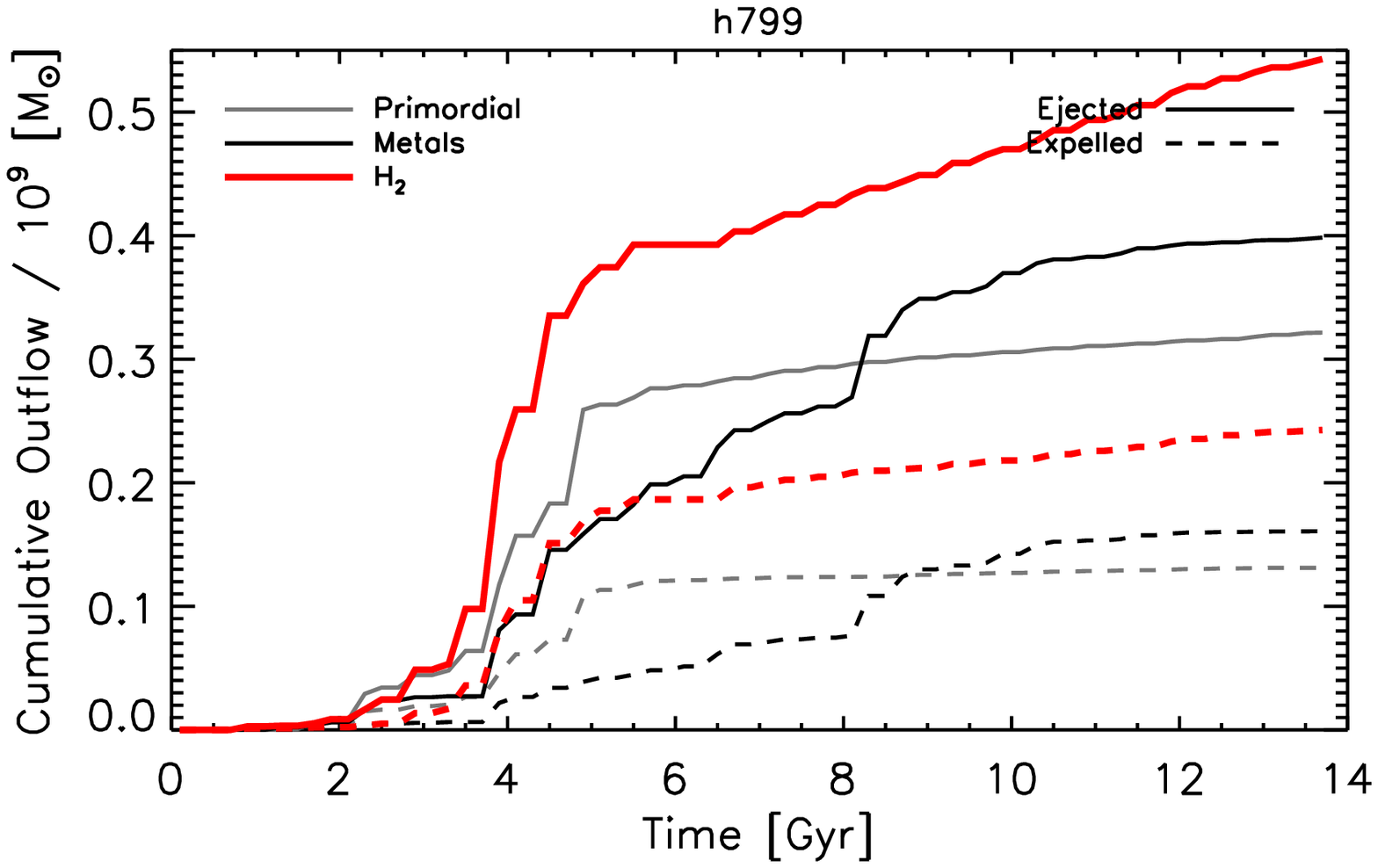}
\includegraphics[width = 0.5\textwidth]{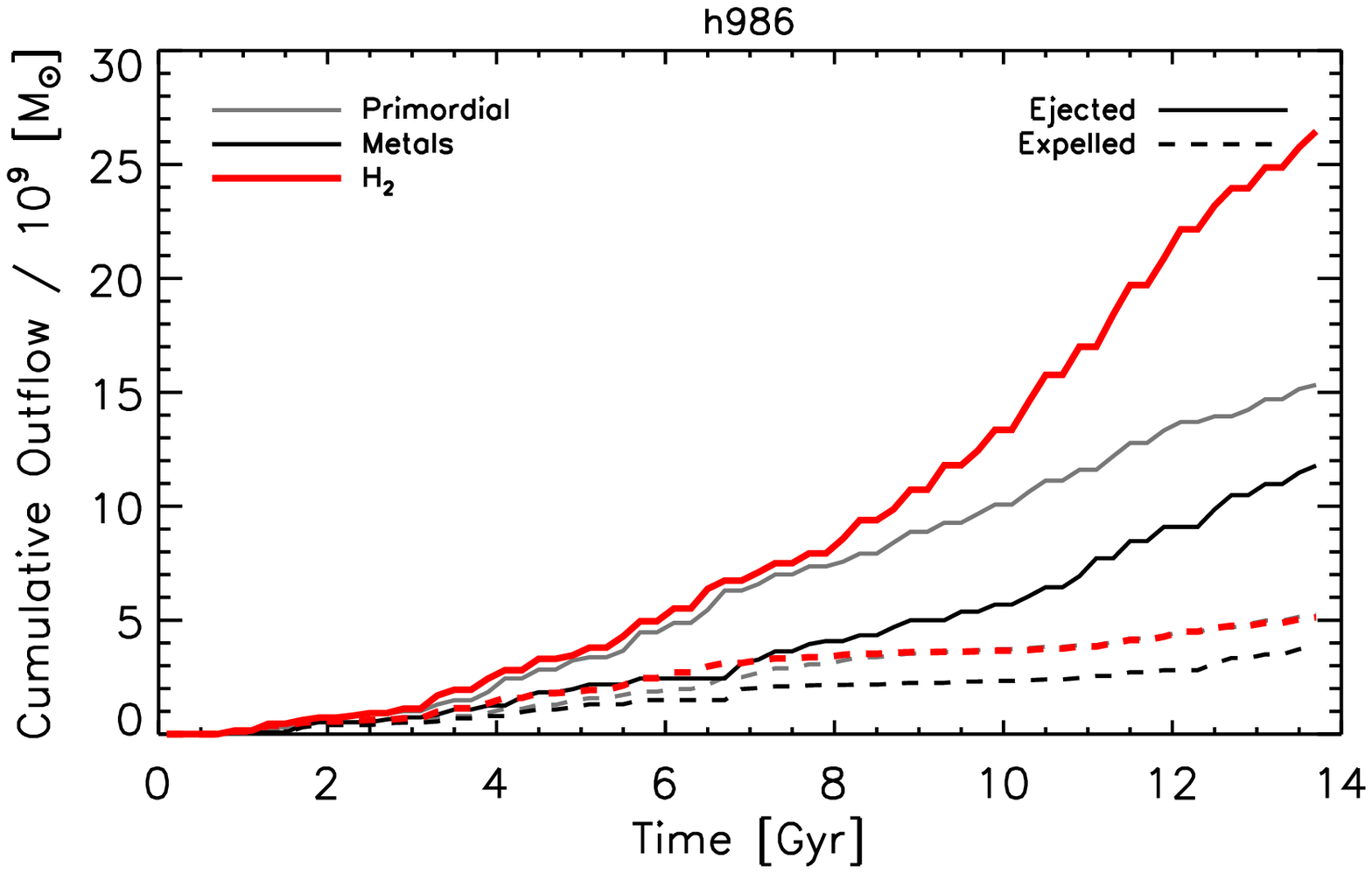}
\end{array}$
\end{center}
\caption[Amount of Gas lost]
{
Cumulative history of gas loss of the representative dwarf (h799) and spiral (h986) galaxies.
Red lines represent simulations with the H$_2$ ISM model; black lines represent simulations with the metal ISM model; grey lines represent simulations with the primordial ISM model.
Solid lines show the mass of gas ejected from the galaxy disk and dashed lines show the mass of gas expelled beyond the virial radius.
In all cases, the greatest amount of gas ejection and expulsion happened in the $\Hmol$ galaxy.
In the spiral galaxy, this was followed by the primordial ISM and then metal line cooling whereas in the dwarf galaxy, a greater amount of gas was lost from the metal ISM model than the primordial ISM model.
}
\label{fig:histfb}
\end{figure*}

The mass of gas lost is shown in Figure~\ref{fig:histfb}.
In this figure, the cumulative histograms of gas ejected from the disk and expelled from the main halo over time are shown for a representative dwarf and spiral galaxy simulated with the three different ISM models.
The effect of the ISM on the spiral galaxies follows the same pattern as the AMD suggests: the primordial and $\Hmol$ ISM galaxy experienced the most gas ejection from feedback, followed by the galaxy with the metal ISM model.
The spiral galaxies simulated with either the primordial or $\Hmol$ models also had substantially reduced central concentrations than the metal ISM model.
This pattern supports the theory that feedback-induced ejection of gas is responsible for changing the central mass distribution of galaxies.
The changes to the AMDs for the dwarf galaxies where much more subtle.  
However, the dwarf with the most gas loss (the $\Hmol$ ISM) is also the one with the smallest amount of low-angular momentum baryons whereas the one with the least gas loss (the primordial ISM) was the one with the greatest amount of low-angular momentum baryons.
We address why different amount of gas loss had smaller effects of the dwarf than the spiral galaxy later on in this section when we compare the mass loading factors and origin of ejected material.

The difference between the ``ejected" and ``expelled" gas represents the fraction of gas particles that either remain in the halo indefinitely or are re-accreted onto the galaxy.
As this paper primarily concerns itself with the distribution of material in the disk, rather than the halo, the ejected material is more relevant to our analysis than the expelled material.
However, the ratio of them to each other is interesting.
The addition of metal line cooling increased the amount of cooling experienced by ejected gas particles in the halo and made them less likely to be completely expelled.
This change manifested itself in the slightly smaller ratios of expelled to ejected material in both the "metals" and "$\Hmol$" ISM models.
Previous research with simulations has also shown galactic winds can enrich the circumgalactic gas \citep{Oppenheimer06, Ford12} and that a substantial amount of gas in galactic winds is re-accreted at a later time \citep{Oppenheimer06, Oppenheimer10}.
Despite our different feedback models and methods for identifying outflows, our simulations appear to be generally consistent with these previous results.

%--------------------------------------------- Mass-Loading ------------------------------------------------------------
\begin{figure*}
\begin{center}
$
\begin{array}{cc}
\includegraphics[width = 0.5\textwidth]{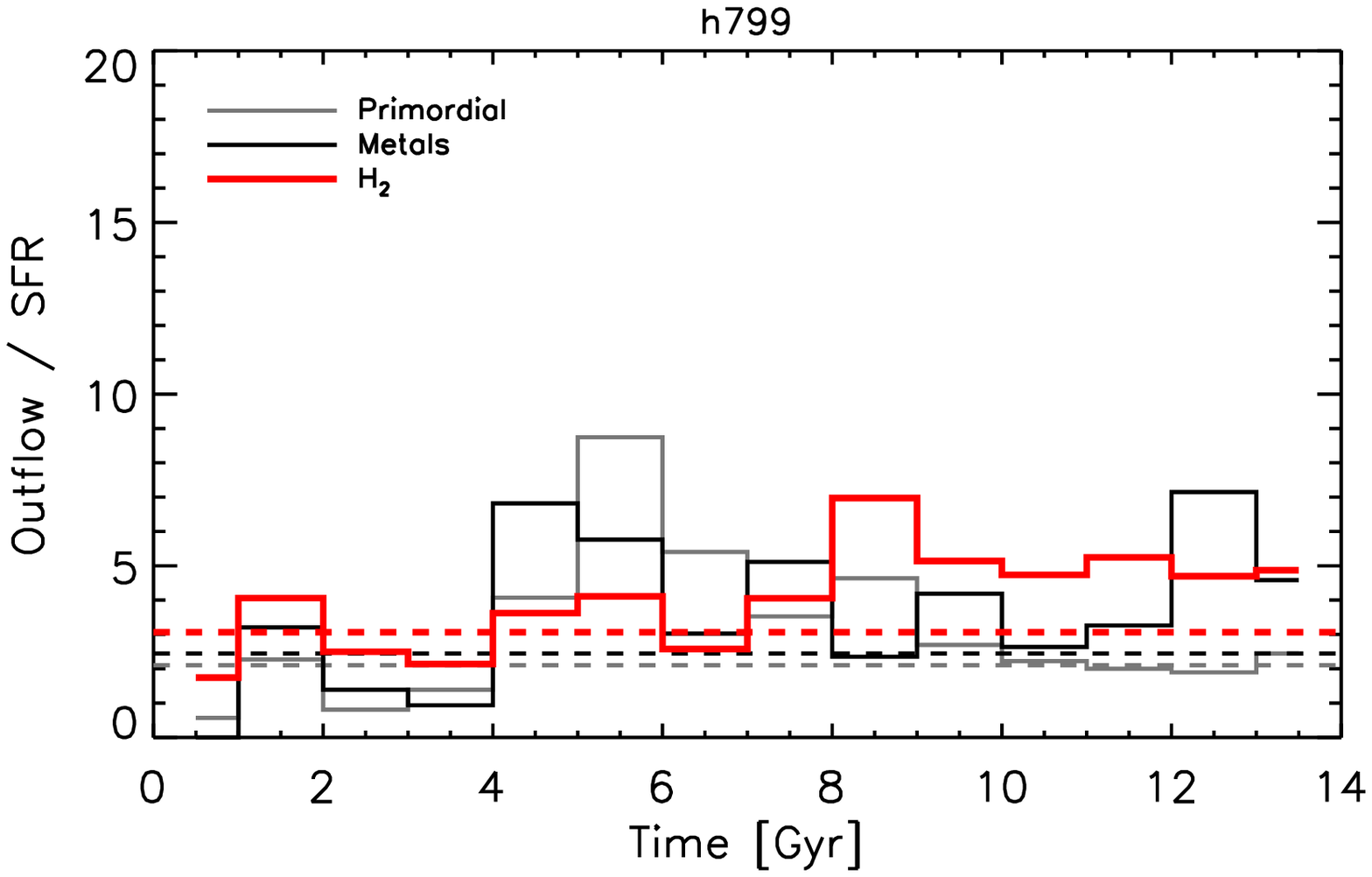}&
\includegraphics[width = 0.5\textwidth]{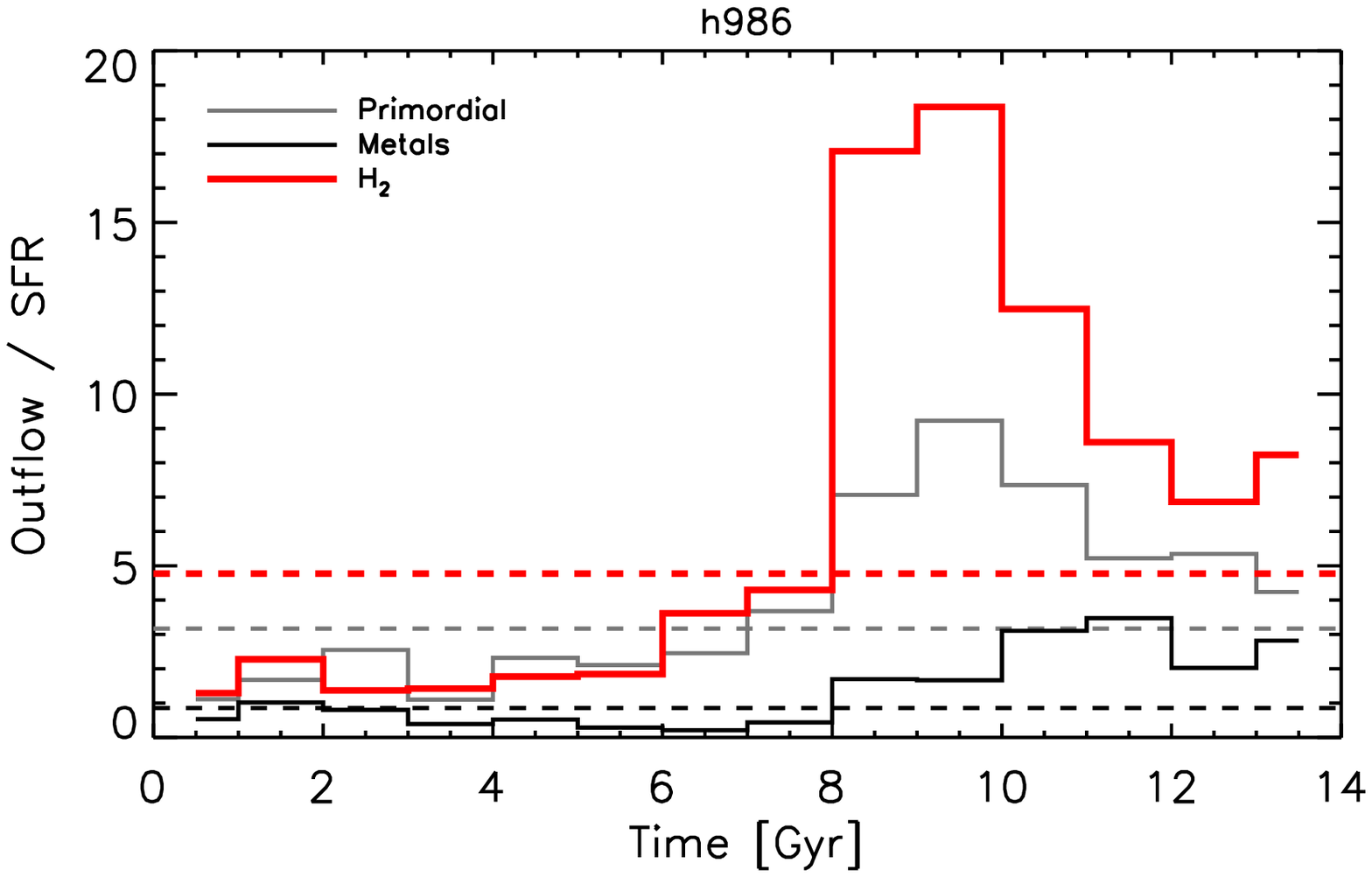}
\end{array}$
\end{center}
\caption
{
Mass loading factor (mass of gas ejected divided by mass of stars formed in the main halo) for a representative dwarf (h799) and spiral (h986) galaxy simulated with the three different ISM models.
The solid lines show the mass loading factor in 500 Myr time bins while the dashed line shows the mass loading factor averaged over the entire history of the main halo.
The red lines represent simulations with the H$_2$ ISM model; the black lines represent simulations with the metal ISM model; the grey lines represent simulations with the primordial ISM model.
The different ISM models had negligible effects on the mass loading factors of the dwarf galaxies.
However, in the spiral galaxy, the $\Hmol$ ISM model had the highest mass loading factors, followed by the primordial ISM model.
}
\label{fig:massload}
\end{figure*}

The total amount of gas ejected from the galaxy disk is a function of the SFR.
Therefore, in order to compare the efficiencies of SNe at ejecting gas in these simulations, we divided the mass of gas ejected by the amount of stars formed within the main halo to arrive at a mass loading factor (Figure~\ref{fig:massload}).
It is clear that in the spiral galaxy both the $\Hmol$ and primordial ISM models resulted in larger mass loading factors than the metal model.
This trend is as expected from the comparison of the mass profiles and bulges and explains the greater amount of mass ejection observed in these galaxies.
In contrast, different ISM models in the dwarf galaxies resulted in very similar mass loading factors, indicating that the different ISM models had much less effect on the structure of the gas.  
The different amounts of gas ejection are, therefore, the result of the different accretion and star formation histories of the main halo.
It is somewhat surprising that the mass loading factors of the spiral galaxy with $\Hmol$ surpass those of the dwarf galaxy. 
More extensive work is needed to examine the relationship between the average mass loading factor and the galaxy mass.
However, preliminary work indicates that there is a range of scatter in the relationship.

\begin{figure*}
\begin{center}
$
\begin{array}{cc}
\includegraphics[width =0.5\textwidth]{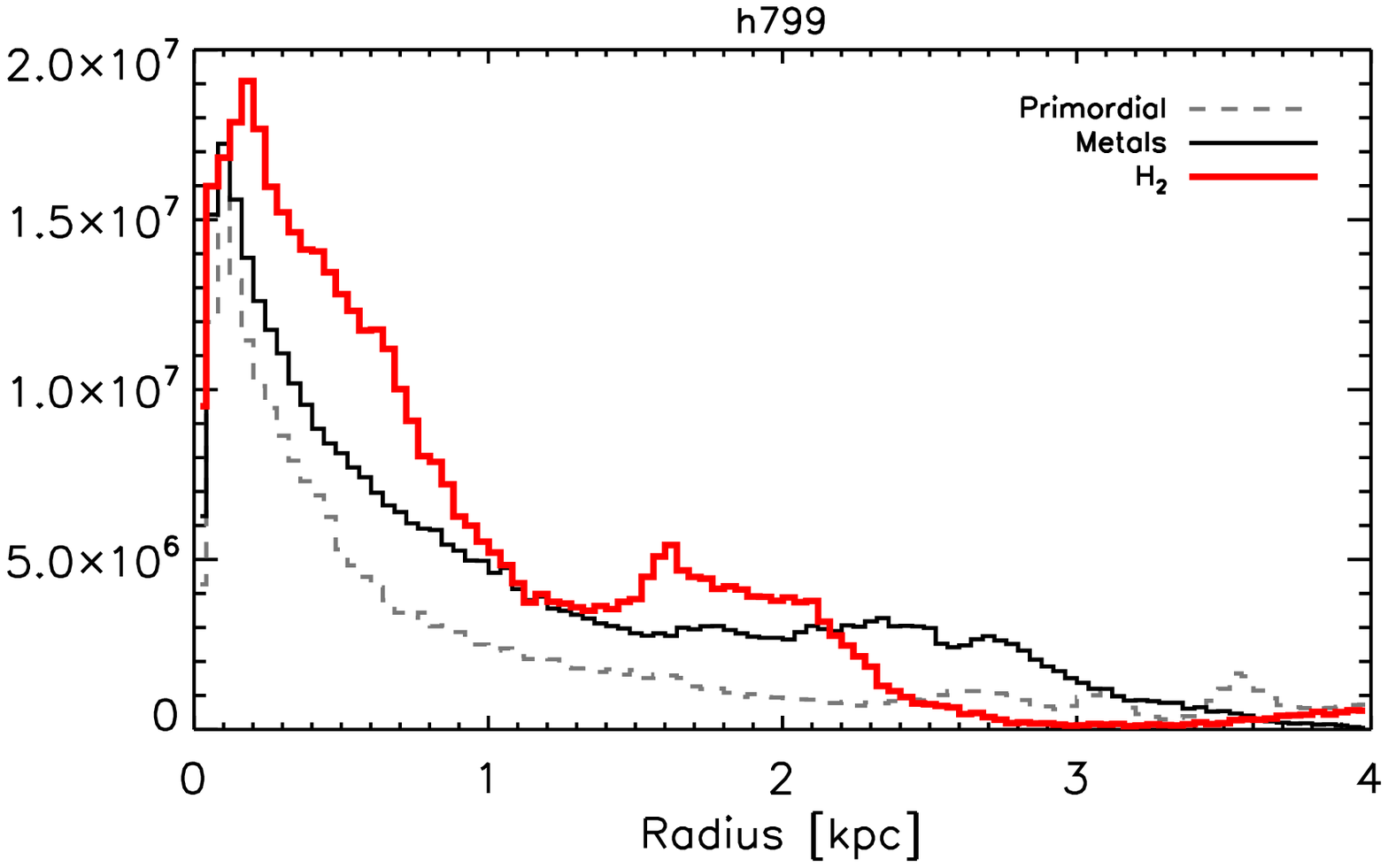}&
\includegraphics[width =0.5\textwidth]{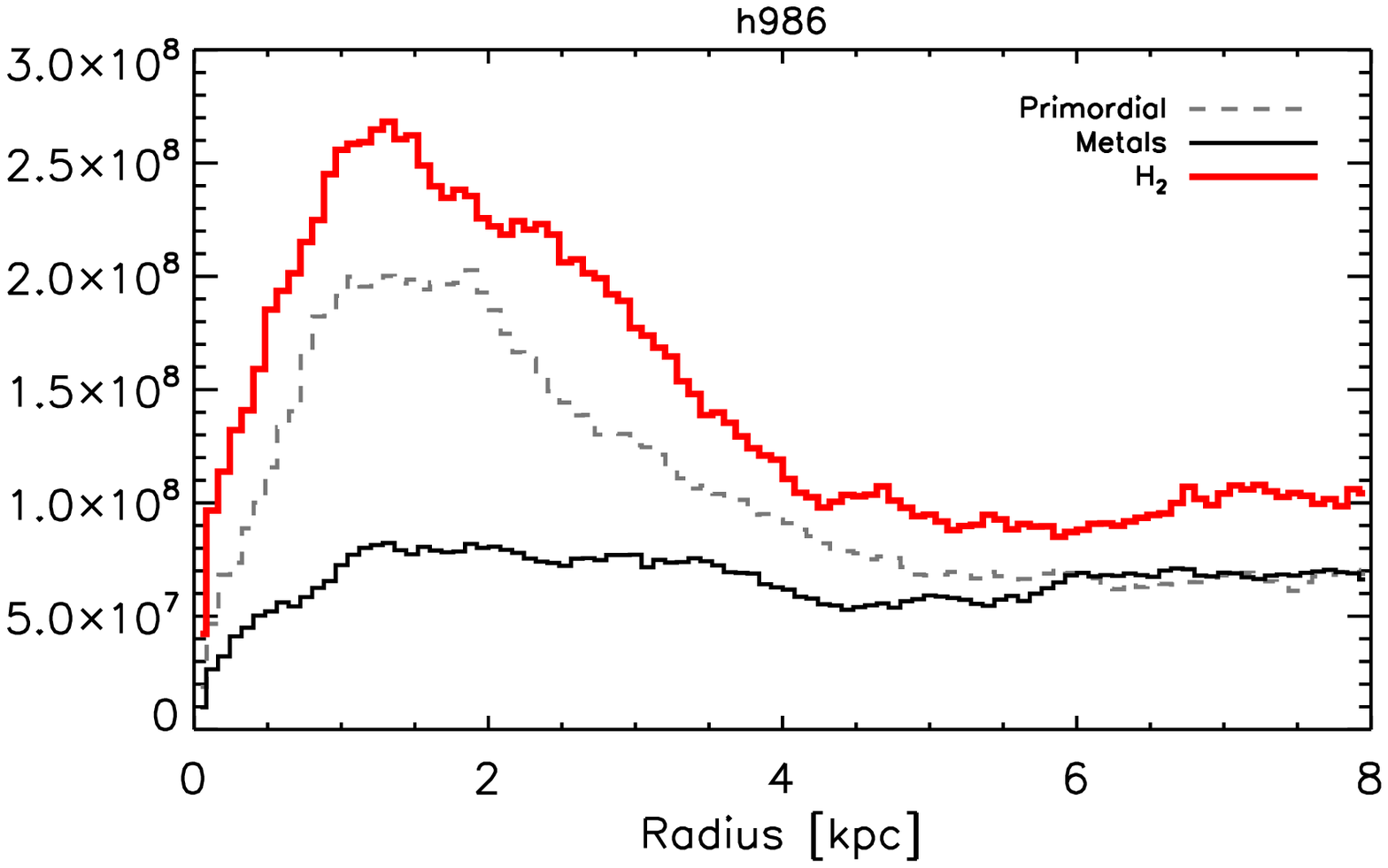}
\end{array}$
\end{center}
\caption[Where gas is being expelled from]
{
Distribution of gas particle radii prior to the gas being ejected from the representative dwarf (h799) and spiral galaxy (h986).
Red lines represent simulations with the H$_2$ ISM model; black lines represent simulations with the metal ISM model; grey lines represent simulations with the primordial ISM model. 
We determined the distribution of radii of gas in the disk before its ejection by selecting for all supernova-heated particles that were in the disk of the galaxy at one timestep and not at the following timestep. 
We then compared the distributions of those radii at the step prior to ejection.
Substantially more mass in the central kiloparsecs is ejected from the galactic disk of the spiral galaxies with the $\Hmol$ and primordial ISM models compared to the metal ISM model.
The distribution of gas ejected from the dwarf galaxies, though, are much more similar.
}
\label{fig:radiusfb}
\end{figure*}

In order to better understand the connection between feedback and central concentration, we examined from where in the disk the gas was originating.
The distribution of the radii of the gas particles prior to being ejected is shown in Figure~\ref{fig:radiusfb} for the representative dwarf and spiral galaxy with the three different ISM models.
To determine the originating radii, we identified the last timestep the gas particle was in the disk prior to being ejected.
The radii in cylindrical coordinates for each gas particle when it was last in the disk was recorded to create the histograms.
The distributions of radii illustrate the importance of gas loss from the center of the galaxy.
For the spiral galaxy, the primordial ISM model and the $\Hmol$ model produce larger amounts of gas loss from the central couple of kiloparsecs, as anticipated from the reduced central masses of these galaxies.
While it is probable that the smaller central stellar masses produced by the primordial ISM model and the $\Hmol$ model are in part due to greater feedback regulation of star formation during mergers (see Figure~\ref{fig:sfh}), this comparison illustrates that feedback also drove centralized mass loss.
In contrast, the distributions of the originating radii for the dwarf galaxy are much more similar, which is consistent with the small changes to the profiles and AMDs.
The slightly larger area from which gas is expelled in the $\Hmol$ model is the result of the larger spatial distribution of star formation.
This greater area of star formation is apparent in the larger optical radii (Table 1) and was previously discussed in \citep{Christensen12}. 

\section{Discussion}\label{sec:discuss}
The correspondence between gas loss and central mass concentration in galaxies illustrates how stellar feedback can reduce the amount of low-angular momentum material in galaxies.
The differences between the ISM models are likely the result of two different competing effects.
On the one hand, greater cooling allows heated gas to more easily cool back onto the disk and insures that gas will require greater amounts of energy to be expelled from the galaxy.
On the other hand, changes to the star forming gas can also lead to a more efficient transfer of feedback energy to gas \citep{Governato10, Guedes11, Governato12}.

The comparatively high efficiency of gas loss from spiral galaxies with the primordial ISM model was the result of the lower cooling rates compared to the other two models.
Gas heated by SNe in this model both starts at higher temperatures ($\geq 10^4$K) and cools more slowly than when metal line cooling is included.
The initially high temperatures of the gas could be seen as similar to the effect of ``pre-heating" the gas by stellar radiation (see \citet{Stinson13})
The result of this decreased cooling in our simulations of spiral galaxies was a greater amount of gas ejected farther from the disks.
When metal line cooling is added, the temperature of the star forming gas decreases and the cooling rates increases.
As a result, greater amounts of energy are required to eject the gas and once ejected, the gas cools and re-accretes onto the disk more quickly.
In our simulations of spiral galaxies, these effects resulted in galaxies with more low-angular momentum material and higher central concentration.
A smaller fraction of the gas remained outside of the disk long enough to be identified as ``ejected" and a smaller fraction was expelled from the galaxy, resulting in smaller mass loading fractions. 
The higher rates of cooling for the halo gas in the metal ISM model more closely replicate the physical Universe but the effects of it must be adjusted for.
Other simulations with metal line cooling have anticipated the effect of the increased cooling on feedback and compensated for it by increasing the amount of energy released per SN \citep[e.g.][]{Pilkington11}. 
A full comparison such as ours, in which the energy released per SN is held constant, is necessary, to find the extent to which the additional cooling affects the matter distribution in simulations.
However, it is possible that future simulations could compensate for the greater amounts of cooling in the metal ISM model through increasing the amount of SN energy or otherwise tuning star formation and feedback parameters.

When $\Hmol$ is added, the inclusion of the dust shielding for both HI and $\Hmol$ and self-shielding of $\Hmol$ results in larger amounts of cold gas.
In our simulations, the formation of this cold, shielded gas greatly increased the clumpiness of the ISM, as was evident when comparing the gas densities of the particles and the power spectrum of the HI surface density.
Furthermore, by linking star formation to the $\Hmol$ abundance, star formation was forced to occur in much denser gas than in either the primordial or metal ISM models.
These changes to the ISM result in greater supernova-driven mass loss in the spiral galaxies, as seen by the increased amount of ejected and expelled material compared to the Metal ISM model.
This increased efficiency happens primarily because the supernova are occurring in a high-density environment that is still relatively low-pressure.
The ability of gas to become molecular and reach colder temperatures at the same density mean that the pressure is lower for a given density of gas.
 As the blastwave model of feedback assumes that supernova cavities last longer in low-pressure and high-density environments \citep{McKee77} more particles have their cooling shut-off for longer time periods and the amount of ejected gas is increased.
Despite the increased amount of ejected gas, a smaller fraction of that ejected gas than in the primordial model was fully expelled from the galaxy because of the metal line cooling of halo gas.
This increased feedback efficiency resulted in decreased central concentration and smaller bulges than with the metal ISM model.

These results raise the question: what aspects of the $\Hmol$ ISM model primarily caused the differences in the simulations?
The first change was the increased amount of cold gas generated by the dust and self-shielding, which resulted in a clumpier ISM.
The second change was forcing of star formation to occur in shielded, dense gas, where the $\Hmol$ abundances were high.
To first order, both of these effects could be reproduced by implementing a dust shielding model for HI and a star formation recipe that was a function of the local shielding.
Such star formation laws that are a function of the shielded gas (as opposed to the $\Hmol$) have been proposed \citep{MacLow10,Krumholz12,Glover11a} as a physical mechanism to explain the observed connection between $\Hmol$ and star formation.
It is possible that the time delay introduced by requiring shielded gas to form $\Hmol$ may result in subtle differences between $\Hmol$-based and shielding-based star formation laws.
In the future, detailed simulations may be able to tease the effects apart.

In contrast to the spiral galaxies, the dwarf galaxies show much less variation in their profiles and their mass loading factors.
The smaller dependency of the dwarf galaxies on the ISM model is the result of their lower metallicity.
Lower metallicities result both in less metal line cooling and less $\Hmol$.
Therefore, the resulting ISMs are more similar across the different models.
The contrast between the dwarf galaxies and spiral galaxies illustrates the importance of testing galaxies of different masses.

Our analysis also demonstrates how the K-S relation, one of the most common metrics for analyzing galaxies, was insufficient for comparing our galaxy models.
We found that our very different ISM models and star formation criteria were all able to produce galaxies that lay along the resolved K-S relation.
Even when measured over 750 pc squares, small-scale variations in the ISM were washed out.
Furthermore, in these simulations feedback is such a strong regulator of star formation that differences between the star formation thresholds and star formation efficiency had a relatively small effect.
We found the gas particle densities and HI surface density power spectrum to provide a more sensitive comparison of the structure of the ISM produced with each of the models.

\section{Conclusions}\label{sec:res5}
We analyzed a set of high-resolution dwarf and spiral galaxies computed with three different ISM models: 1) primordial cooling only, 2) metal line cooling and 3) metal line cooling with gas shielding, non-equilibrium $\Hmol$ formation, and an $\Hmol$-based star formation recipe.
These galaxies were simulated in a cosmological context to a redshift of zero.
In order to make our results most comparable to observational data, we analyzed the stellar distribution and mass and the density structure of the ISM at $z = 0$ in an observationally motivated fashion through the creation of simulated optical images and HI velocity cubes.
We determined the effect of different ISM models on the efficiency of feedback in redistributing matter throughout the galaxy and we successfully created spiral galaxies with reduced amounts of low-angular momentum material.
We found the following results for our suite of simulations:

\begin{itemize}
\item The different ISM models resulted in clear changes to the centers of the spiral galaxies.  Spiral galaxies simulated with both the primordial and $\Hmol$ ISM model had reduced masses within the central 1kpc compared to the metal ISM model, which was apparent in their realistic non-centrally peaked rotation curves, less concentrated DM, and reduced mass of low-angular momentum material.

\item Increasing amounts of cold gas resulted in clumpier ISMs.  The effect was particularly strong for the spiral galaxies with $\Hmol$.  These galaxies had shallower power structures and increased amounts of high density gas.  The ability of gas to collect in high-density clumps at larger radii resulted in more extended star formation at z = 0{, although the structure of the stellar disks was similar across models.}

\item  All simulated galaxies fell along the observed K-S relation (resolved to bins of 750$^2$ pc$^2$).  However, the individual star forming particles had very different density and temperature distributions for the different models.  In particular, in the $\Hmol$ ISM model stars formed from colder and denser gas because of the presence of shielding and the use of an $\Hmol$-based star formation law.

\item The star formation histories of the spiral galaxies with the metal ISM model showed increased star formation around a redshift of one compared to the other two models.  However, there is little difference between the star formation histories for the spiral galaxies prior to 3 Gyrs.  This similarity at high redshift suggests that the reduced stellar masses resulting from the $\Hmol$ ISM model were not the result of reduced star formation efficiency in low-metallicity gas early in the galaxies's histories.

\item The spiral galaxies simulated with the primordial and $\Hmol$ ISM models had greater amounts of gas ejected beyond the disk by SNe and larger mass loading factors.  This increased efficiency of feedback was likely responsible for the changes to the central kpc of the galaxies.  The galaxies with the metal and $\Hmol$ ISM models had smaller fractions of their ejected gas escape beyond the virial radius because of increased cooling in the halo from metal lines.

\item The dwarf galaxies simulated with different ISM models had relatively similar structures, including rotation curves and stellar profiles, and similar mass loading factors.  The low metallicities of these galaxies meant that there was less potential for metal line cooling or dust shielding to affect the structure of the ISM.

\end{itemize}

The fact that both decreased cooling through lack of metal line cooling and increased cold gas from the shielding in our $\Hmol$-model resulted in decreased bulges and increased feedback efficiency implies that multiple processes are at work in determining the mass of bulge stars formed.
Compared to galaxies formed with only primordial cooling,  the additional cooling in the metal ISM model means that more feedback energy is required to expel particles.
When using feedback parameters tuned for only primordial cooling, spiral galaxies simulated with metal-line cooling have more-cuspy DM profiles, greater amounts of low-angular momentum gas, centrally-peaked rotation curves, and bigger and more concentrated bulges.
With the addition of our $\Hmol$ model, the efficiency of feedback once again increased.
In this model, stars formed out of and supernova occurred in dense and cold molecular gas.
The combination of higher densities and lower pressure insured that cooling in the blastwave model was turned off for a longer period of time and increased likelihood of ejecting gas particles from the disk.
The increased efficiency of feedback resulted in spiral galaxies with smaller central DM concentrations and decreased amounts of low-angular momentum material, which led to more realistic rising rotation curves and smaller central masses.
{\em This final model is able to reduce the central concentration of galaxies by including the metal line cooling necessary for modeling the CGM and the re-accretion of gas onto the disk and by including a model for shielded, molecular gas and the star formation that raised the feedback efficiency enough to produce the observed centers of galaxies.}

Until recently, most simulation focused on replicating the disks of galaxies.
As the disks of simulated galaxies have become more realistic, however, the production of bulges with appropriate masses and concentrations has remained a difficult problem.
The large amount of computing time required to simulate the formation of a spiral galaxy with the necessary sub-kpc resolution means that very few simulations have been able study the bulges of galaxies formed in a cosmological context \citep{Agertz10,Guedes11,Okamoto12}.
Nevertheless, the sensitivity of central 1 kpc in galaxies to angular momentum loss, make them a prime area of research.
In a future work, we will examine how the reduction of the central mass in galaxies through feedback can result in bulges with realistic structure.

Forming realistic centers of galaxies requires reducing the amount of low-angular momentum material in galaxies.
Together, these simulations reinforce the idea that stellar feedback is an effective way to remove low-angular momentum material from galaxies and to shape the central DM distribution.
They also demonstrate that stellar feedback models are inexorably linked to ISM models in simulations.
In order to create realistic galaxies, models of feedback, ISM and star formation must work together in a realistic manner.

\section*{Acknowledgments}
The authors would like to thank the anonymous referee for the extremely suggestions.
We also thank Greg Stinson for the useful discussions.
Resources supporting this work were provided by the NASA High-End Computing (HEC) Program through the NASA Advanced Supercomputing (NAS) Division at Ames Research Center.
Further simulations were computed at the Texas Supercomputing Center.
CC acknowledges support from NSF grants AST-0908499 and AST-1009452.
FG acknowledges support from HST GO-1125, NSF AST-0908499.
TQ acknowledges support from NSF grant  AST-0908499.
AB acknowledges support from The Grainger Foundation.
We made use of pynbody (http://code.google.com/p/pynbody) in our analysis for this paper.

\bibliographystyle{apj}
\bibliography{ISMBulges.bib}

\end{document}